\documentclass[final,5p,times,onecolumn,authoryear]{elsarticle}
\usepackage{xcolor}
\usepackage{aas_macros}
\usepackage{amssymb}
\usepackage{lipsum}
\usepackage{amsmath}

\journal{New Astronomy}

\begin{document}

\begin{frontmatter}

\title{Prospects for high-resolution probes of galaxy dynamics tracing background cosmology in MaNGA}

\author[first]{Gyeong-Min Lee}
\author[second]{Maurice H.P.M. van Putten\corref{cor1}\fnref{label2}\fnref{label3}}
  \ead{mvp@sejong.ac.kr}
\fntext[label3]{INAF-OAS Bologna via P. Gobetti 101 I-40129 Bologna Italy, Italy}
\affiliation[first]{organization={Physics and Astronomy, Sejong University}, addressline={209 Neungdong-ro}, postcode={05006},
            city={Seoul},country={South Korea}
         }

\begin{abstract}
Large-$N$ galaxy surveys offer unprecedented opportunities to probe weak gravitation in galaxy dynamics that may contain correlations tracing background cosmology. Of particular interest is the potential of finite sensitivities to the background de Sitter scale of acceleration $a_{dS}=cH$, where $H$ is the Hubble parameter and $c$ is the velocity of light.
At sufficiently large $N$, this is possibly probed by ensemble-averaged ("stacked") rotation curves (RCs) at 
resolutions on par with present estimates of the Hubble parameter $H_0$.
Here, we consider the prospect for 
studies using the large $N$ {\it Mapping Nearby Galaxies at Apache Point Observatory} MaNGA at APO survey. 
In a first and preliminary step, we consider unbiased control of sub-sample size by consistency in the three Position Angles, $\theta$ , from photometry and velocity fields of gas and stars by spectroscopy within $30^\circ$.
In sub-samples of size $N=N_i(\theta)$, the scatter in-the-mean $\sigma/\sqrt{N}$ is found to reach one percent levels, differentiated over inclination angle $i$ and $\theta$. 
In regular propagation of uncertainties, this scatter contributes $\kappa\sigma/\sqrt{N}$ to the standard error in-the-mean to the observable, where $\kappa$ is determined by the choice of observables.
As a lower bound to scatter in stacked RCs, MaNGA hereby appears promising for high-resolution analysis of sensitivity to RCs to background cosmology, notably across a sharp $C^0$-transition \citep{van18} of Newtonian to anomalous dynamics across $a_{dS}$ and, further out, the baryonic Tully-Fisher relation. 
In turn, these markers provide a novel measurement of cosmological parameters.
\end{abstract}




\end{frontmatter}

\onecolumn 

\section{Introduction}

Galaxy surveys offer avenues to probe weak gravitation pioneered in spiral galaxies by \cite{tul77}. A dramatic extension is the discovery of galaxy formation at cosmic dawn by the {\em James Webb Space Telescope} (JWST) \citep[e.g.][]{van24}. 
Modern large-$N$ surveys offer radically new opportunities for high-resolution studies hereof using (sub-)groups of galaxies  with relatively homogeneous statistical properties, e.g., observed by dynamics of gas and stars and model fits. For sufficiently large $N$, this may further allow for binning over different redshifts.

A key question is the feasibility of high-resolution studies with uncertainties on par with modern precision cosmology. 
High-resolution observations of the evolving Universe on galactic and cosmological scales is a hallmark of modern cosmology with the aim to map the Hubble expansion and structure formation in the Universe across all redshifts \citep{rie22,tri24}.

This development is gradually bringing into view significant departures from the concordance model $\Lambda$CDM \citep{div21}, which assumes a constant vacuum energy density $\Lambda$ and an asymptotic de Sitter cosmology in the distant future.
Notably, the expansion history at early and late times reveals a tension in the Hubble constant, in {\em Planck} $\Lambda$CDM-analysis of the {\it Cosmic Microwave Background } (CMB) \citep{tri24} and, independently, in the Local Distance Ladder (LDL) \citep{rie22}. 
This $H_0$-tension has reached a significance of $5\sigma$ by resolving $H_0$ at statistical uncertainties at about the one percent \citep{rie22,div21}. Hubble expansion consistent with early and late-time cosmology may point to a dynamic dark energy \citep{van25}.

In light of this, it is critical to also test $\Lambda$CDM for its predictions on sub-horizon scales, in the growth of large scale structure formation \citep[e.g.][]{2021CmPhy...4..123J}, the dynamics of galaxy clusters \citep{1937ApJ....86..217Z,per22,nat24,hua24}, 
and galaxies dynamics \citep{1970ApJ...159..379R,tul77,mil83,fam12,lel16,mcg16,san02,nes23}.

This suggests the need for one-percent precision studies of galaxy dynamics - a standard error-in-the mean of about 1\% in large-$N$ ensemble-averaged observables. Such can  then be used to study potentially subtle correlations of galaxy dynamics with the cosmological background, hitherto hinted at using relatively small-$N$ surveys \citep{van18}. 
To study this potential model-independently in some generality, we here set out to quantify consistency in photometric-spectroscopic data of spiral galaxies by some canonical statistics to be specified further below.

Photometric observations provide extensive morphological classification of {galaxy dynamics} \citep{1927Obs....50..276H,mas21,ndu23,med24}. 
Spectral energy distributions (SED) by multi-wavelength observations further provide detailed probes of a variety of properties {such as the} Initial Mass Function (IMF) of stars and star formation rates (SFR) and mass-to-light ratios $M/L$
\citep{sal55,bel01,bri04,mad14,zha24}, essential to the direct measurement of baryonic matter content.
For spiral galaxies, the well-established Tully-Fisher relation (TFR, \cite{tul77}) {stands out, revealing} anomalous galaxy dynamics based on baryonic matter content alone (bTFR, \cite{mcg00,mcg12}).

In late-time cosmology, the physical origin of bTFR is ambiguous, however, arising from dark matter conform $\Lambda$CDM {\em or} non-Newtonian dynamics in departure thereof. In contrast, $\Lambda$CDM is up-ended at cosmic dawn by JWST `Impossible galaxies', when dark matter is already maximal \citep{van24}. 
The bTFR and the ultra-high redshift JWST galaxy may have a potentially unified origin in a finite sensitivity of galaxy dynamics, including galaxy formation at cosmic dawn, to the background de Sitter scale of acceleration \citep{van17a,van24,van24b}.
\begin{eqnarray}
a_{dS}=cH, 
\label{EQN_adS}
\end{eqnarray}
where $c$ is the velocity of light and $H$ is the Hubble parameter. 

Detailed studies of the bTFR by modern large $N$ galaxy surveys resolve galaxy rotation curves by high-resolution spectroscopy in the footsteps of the first pioneering Doppler shift observations \citep{1915PA.....23...21S}. 
This is realized by long-slit observations, providing one-dimensional spatially resolved spectra, 
and two-dimensional maps by Integral Field spectroscopy (IFS) of individual galaxies \citep{2003MNRAS.342..345C,lu23,lu24}. 

The combined photometric-spectroscopic observations produce maps of galaxy dynamics at high resolution, providing a starting point for comprehensive and accurate probes of galaxy dynamics. Potentially, this brings into view the transition of Newtonian dynamics to 
anomalous galaxy dynamics further out in galactic disks, that forms the basis of the bTFR.
At sufficiently high resolution, these maps may identify a finite sensitivity to aforementioned $a_{dS}$. 
Evidence for this sensitivity is found in ensemble-averaged ("stacked") rotation curves (RC) \citep{van17a,van17b,van24} of SPARC \citep{lel16,mcg16}. 
ere, we find a $6\,\sigma$ departure from $\Lambda$CDM galaxy models in a sharp $C^0$-transition across $a_{dS}$ \citep{van18}, distinct from the much smaller scale of acceleration $a_0$ that parametrizes bTFR of the asymptotic behavior at large radii \citep{mil83,mcg12}.

Probes of galaxy dynamics in the {\em Mapping Nearby Galaxies at APO} (MaNGA) survey carry a cumulative uncertainty from instrumental observations, data-processing to {stacking}, and interpretation with potential contributions from additional sources of data.
Already, the intrinsic diversity of individual galaxies introduces noise - intrinsic scatter - with additional contributions from a number of inherent observational challenges (e.g. instrument, atmospheric condition, and reducing raw data and sampling). 
The cumulative result defines the total uncertainty budget.
In the interpretation of data, further possibly systematic uncertainties may derive from model assumptions, 
e.g., when involving Mass-to-Light (M/L) ratios based on stellar population synthesis (SPS) and star formation history (SFH) \citep{Conroy13,Zibetti09}.

In this work, we consider a first step in this direction: the prospect of MaNGA to probing sensitivity to $a_{dS}$ in stacked RCs consistent with the approximately one-percent uncertainty of modern estimates of $H_0$. 
{In a necessary but preliminary assessment, we focus on statistical uncertainty - noise from scatter - in sub-samples of varying size $N$. An unbiased control of $N$ is given by scatter in-the-mean $\sigma/\sqrt{N}$ in the three Position Angles (PAs) obtained from photometric images and gas and stellar velocity fields.
In a regular propagation of uncertainty (below), 
$\sigma/\sqrt{N}$ is expected to set a minimum 
to the total uncertainty budget in stacked RCs. 
{Aiming to probe} galaxy dynamics at uncertainties on par with observations of $H_0$, we set out to determine
that MaNGA is sufficiently large to allow 
$\sigma/\sqrt{N}$ to reach about one percent.}

MaNGA data release 17 \citep{2015ApJ...798....7B} covers the largest number of galaxies with IFU observations to date. 
For any ensemble-averaged observable, the scatter $\sigma$ in the three PAs of photometric images and the velocity fields of gas and stars by spectroscopy conceivably defines a significant contribution to the standard error in the mean $S_E$. {More} generally, we have \citep[e.g.][]{ku66,hib23,2021ApJ...912...41S}
\begin{eqnarray}
S_E = \frac{\sqrt{\kappa^2\sigma^2+\cdots}}{\sqrt{N}}
\label{EQN_S}
\end{eqnarray}
for a sample size $N$, where $\kappa$ is a constant of order unity assuming regular propagation of uncertainty - the scatter in-the-mean $\sigma/\sqrt{N}$ - and the dots refer to contributions from additional and statistically independent sources of noise in the data. 
To exemplify, of particular interest is the accurate measurement of the bTFR \citep{mcg12} and, 
deeper within galaxies, the location $a_{dS}$ of the $C^0$-transition of Newtonian to anomalous gravitation - in SPARC \citep{van18}, awaiting confirmation in an independent survey such as MaNGA.
 
In \S2, we present our MaNGA sample selection of spiral galaxies. 
In \S3, we present our unbiased parameter controlling the size $N=N_i(\theta)$ of sub-samples by bounding  consistency $\theta$ in the three PAs in MaNGA, binned over various inclination angles $i$. 
Scatter in-the-mean $\sigma/\sqrt{N}$ in (\ref{EQN_S}) as a function of $\theta$ is presented in \S3, 
showing the desired control down to about one percent.
In \S4, we discuss an outlook on probing sensitivity of galaxy dynamics to $a_{dS}$ in stacked MaNGA RCs and its applications to cosmological parameter estimation.

\section{Selection of MaNGA spiral galaxies}

MaNGA is a part of the Fourth-Generation {\em Sloan Digital Sky Survey} (SDSS-IV, \citep{2017AJ....154...28B}). 
It provides spatially resolved spectral measurements of about ~10,000 galaxies. 
The primary sample provides radial coverage over approximately 1.5 half-light radii ($R_e$), included in 67$\%$ of MaNGA. 
The secondary sample extends radial coverage to 2.5\,$R_e$, included in the remaining $33\%$.
The optical wavelengths of BOSS Spectrographs in MaNGA span $0.36\,\mu \lesssim \lambda \lesssim 1.0\,\mu$ 
with a resolution 
$R \sim 2000$ \citep{2019AJ....158..231W}. 
The mean redshift of MaNGA galaxies is approximately 0.03 with a roughly flat stellar mass distribution, $M_{\ast} \gtrsim 10^9\,M_{\odot}$. 

The MaNGA data release includes almost every level of data reduction. 
Starting from raw data, a comprehensive 3-D data reduction pipeline (DRP) extends to a data-analysis pipeline(DAP). 
DRP output is flatfield-subtracted, sky-subtracted, position-calibrated, flux-calibrated, sampled to a common wavelength grid, and so on. 
DAP provides DRP assessments, spatial binning, stellar-continuum modeling, emission-line measurements, spectral measurement and more \citep{2019AJ....158..231W}. 

DAP includes three different binning schemes in DAP output:
\begin{itemize}
\item 
Individual spaxel binning (SPX);
\item 
Spaxel binning with $S/N \sim 10$ by Voronoi (VOR10) \citep{2003MNRAS.342..345C};
\item 
Same as VOR10 binning to stellar continuum analysis for stellar kinematics, and individual binning for emission-line measurements (HYB10). 
\end{itemize}
In stellar continuum analysis providing stellar kinematics, hierarchically clustered MILES templates are used (MILESHC) for all the data. 
In continuum+emission-line modeling, DAP includes two types: simple stellar population models (MASTARSSP) derived from MaStar (MaNGA Stellar Library \citep{2019ApJ...883..175Y}) and hierarchically clustered templates (MASTARHC2) from MaStar.
With the above-mentioned schemes, four DAPTYPE of data are provided. Among these, we select the DAP output HYB10-MILESHC-MASTARSSP combination, 
as HYB10 data is expected to provide the best spatial resolution for emission-line analysis.

Focusing on galaxy rotation curves, we select the MaNGA {\it Deep Learning Morphology Value Added Catalog} (MDLM-VAC-DR17) to select late-type galaxies \citep{2022MNRAS.509.4024D}. 
This catalog provides deep-learning-based morphological classification for MaNGA samples. MDLM-VAC-DR17 used SDSS-DR7 RGB-cutout as input and trained a convolutional
neural network (CNN)  with this input for each classification task. 
Further detailed methodology is provided in \citep{2018MNRAS.476.3661D}. 

In the {MDLM-VAC-DR17} classification, positive T-Type corresponds to late-type galaxies (LTGs), and 
negative T-Type corresponds to early-type galaxies (ETGs). 
According to T-Type, galaxies are subdivided into sub-classes through additional training (CNN) when needed. 
Following the above, {MDLM-VAC-DR17} provides \texttt{P\_LTG}, \texttt{P\_S}0, \texttt{P\_edge-on}, \texttt{P\_bar},
representing the respective probabilities.
T-Type and \texttt{P\_S}0 combined present the three broad categories E, S0, and S. 
It includes a sub-division for S galaxies into S1 and S2 according to the T-Type. 
After visual inspection, they report the result in Visual Class (VC = 1,2,3, and 0 for E, S0, S/Irr and unclassifiable, respectively), 
and Visual Flag (VF=0 for reliable and 1 for uncertain).
Our initial selection of MaNGA sample contains 5125 galaxies according to $\texttt{P\_LTG} > 0.5$, 
positive T-Type, VC\,=\,3 and VF\,=\,0, recommended for spiral galaxies in MDLM-VAC-DR17.

We downselect according to the MaNGA DAP quality flag 
{\it bitmask}, removing 713 galaxies satisfying either of the following: 
(1) identified foreground stars, 
(2) NSA (NASA-Sloan Atlas) redshift inconsistency, 
(3) no emission line, (4) ppxf fit failure, 
(5) Voronoi binning failure, 
(6) invalid input geometry, 
(7) critical failure in DRP, 
(8) critical failure in DAP, or 
(9) critical failure in DRP or DAP.
Table 1 shows the remaining sample of $N=4412$ galaxies. Table 1 is the starting point of our analysis,  
differentiating our sample in bins of inclination angle $i$ by 10$^\circ$ intervals.  
The inclination angles $i$ of the galaxies are calculated from axis-ratios $b/a$, evaluated at 90\% light-radius provided by NASA-Sloan Atlas (NSA) catalog.
We next downloaded above-mentioned DAPTYPE data 'HYB10-MILESHC-MASTARSSP' and 'HYB10-MILESCH-MASTARHC2' from MaNGA for our sample of $N=4412$ galaxies, except for one (MaNGA Plate-IFU 8626-9102, in the bin $50^\circ- 60^\circ$) that appears unavailable from the MaNGA server. This leaves a total count of our MaNGA spiral galaxy selection of \textbf{$N=4411$.}

\begin{figure}
\center{
\begin{tabular}{cc}
\hspace{0.06\linewidth}
\includegraphics[width=0.3\linewidth]{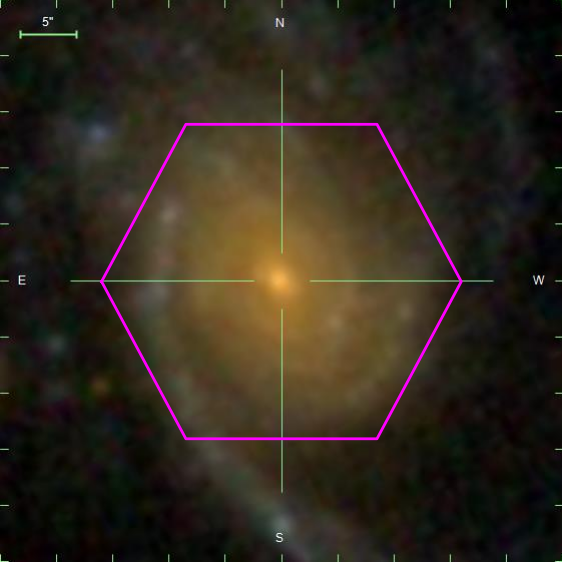} 
\hspace{0.06\linewidth}
\includegraphics[width=0.47\linewidth]{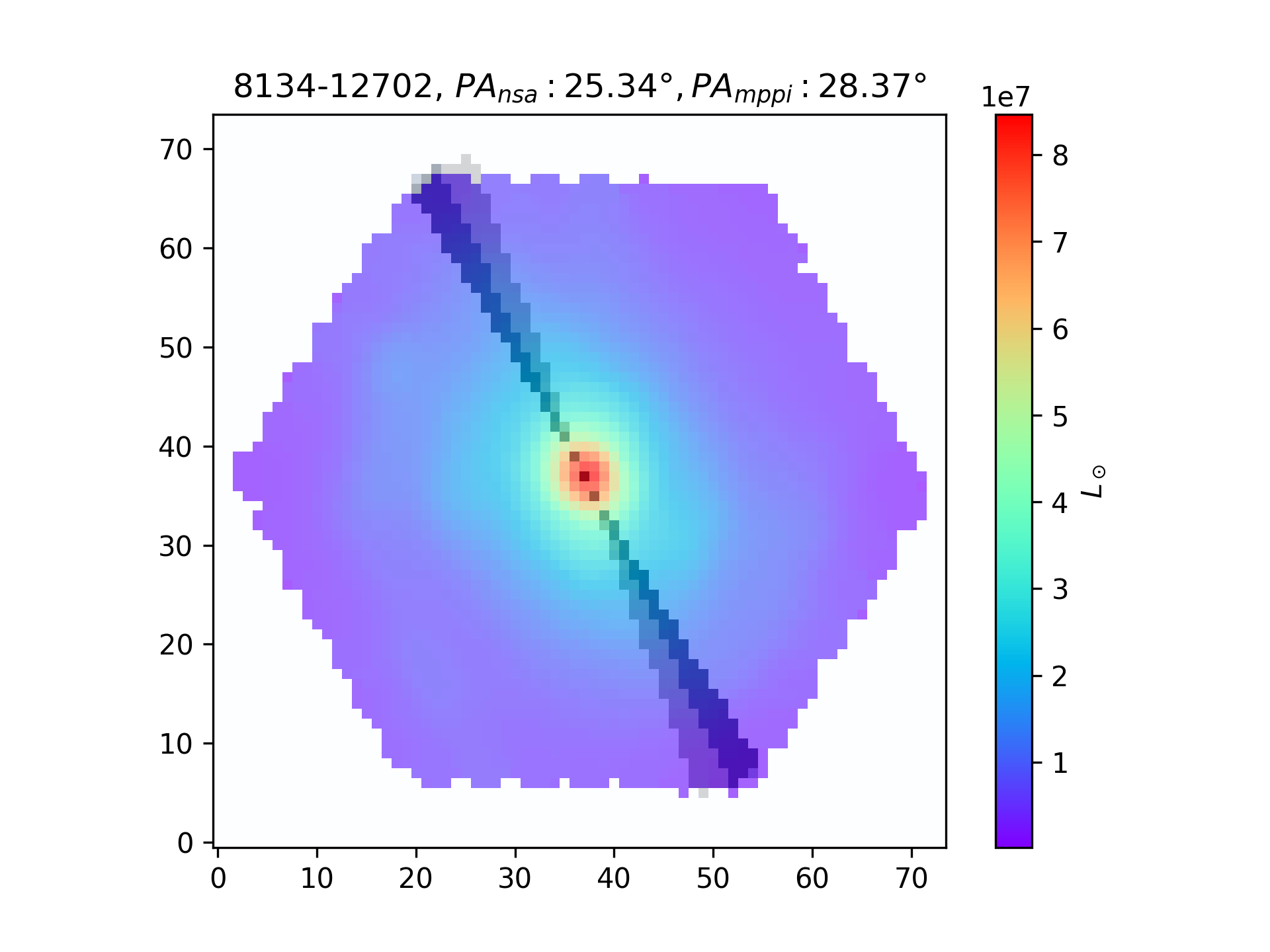}\\ 
\includegraphics[width=0.45\linewidth]{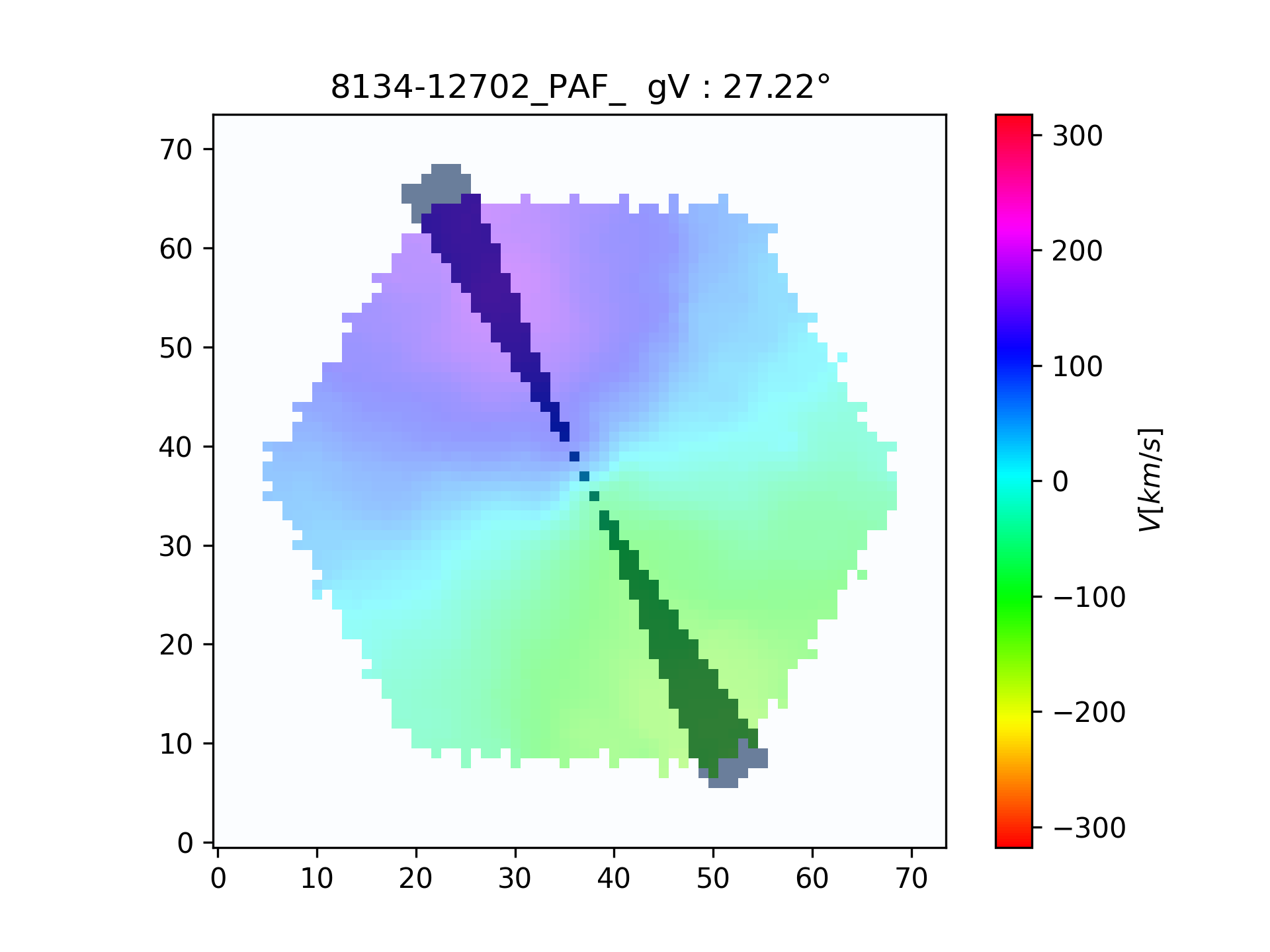} 
\includegraphics[width=0.45\linewidth]{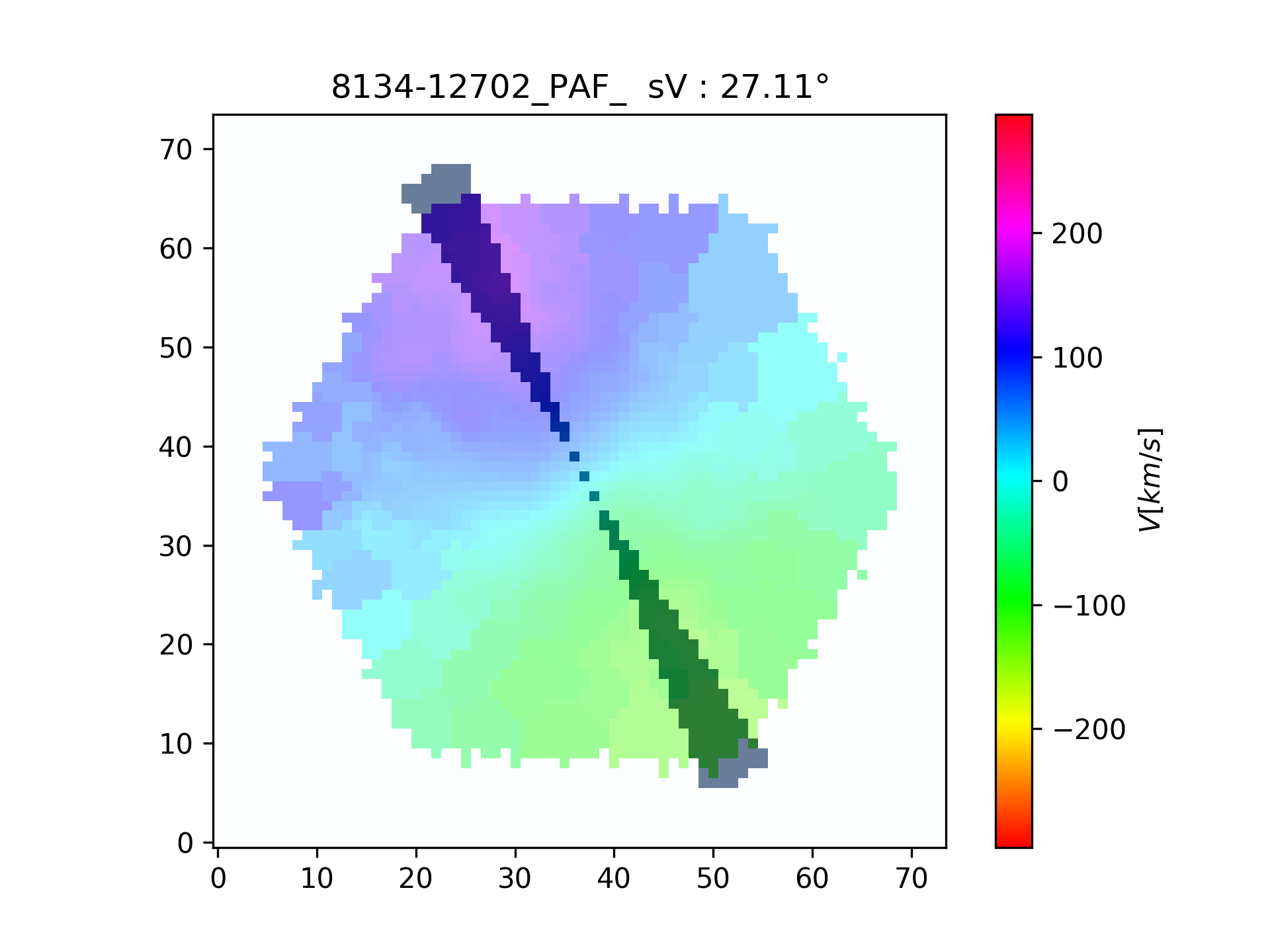} 
\end{tabular}
}
\caption{Photometric and spectroscopic data of a MaNGA galaxy, 8134-12702
MaNGA galaxy 8134-12702. Top left: SDSS photometric image. Top right: MaNGA i-band image overplotted with position angles from NSA catalog and MPP VAC with $\mbox{PA}_{ph} = 25.34^\circ$ and $\mbox{PA}_{ph} = 28.37^\circ$ respectively. Bottom :  Gas velocity map and stellar velocity map from right to the left respectively with small-angle pencil beams (grey shade) show ${\rm PA}_g=27.22^\circ$ and ${\rm PA}_s=27.11^\circ$ in least-squares sine-fits to the velocity field over an opening angle of 12$^\circ$.
}
\label{fig:phot_vel}
\end{figure}

\section{Statistics of consistency in Position Angles}

In a high-resolution analysis of ensemble-averaged data, statistical uncertainty is governed by errors in-the-mean controlled by sample size $N$ for a given scatter $\sigma$ in the data. This consideration does not include systematic errors, that typically is case-specific. 
In regular propagation of uncertainties, contributions to a standard error in-the-mean $S_E$ satisfy canonical scaling with $\sigma/\sqrt{N}$,
the constant of proportionality $\kappa$ highlighted in (\ref{EQN_S}) generally depending on the choice of observable.
For our MaNGA sub-sample of spiral galaxies (\S2), we expect a potentially significant contribution from scatter in the three PAs of photometry and
the velocity fields of gas and stars from spectroscopy, assuming regular propagation of uncertainty quantified by $\kappa$ in (\ref{EQN_S}).
A full account of this contribution to the total noise budget requires measuring $\kappa$ in (\ref{EQN_S}) by
varying the control parameter $\theta$, separate from other (statistically independent) sources of noise in the data. 
To this end, we proceed as follows.

Consistency of the PAs in the velocity fields of gas and stars is defined by the difference 
\begin{eqnarray}
\Delta_1 {\rm PA} = {\rm PA}_g - {\rm PA}_s.
\label{EQN_D1}
\end{eqnarray}

Fig. 1 shows an example of velocity fields of gas and stars. We analyze these data in a polar coordinate system $(r,\varphi)$ with its origin at the galaxy center provided by MaNGA. 
Using best-fit by the least-squares method, we derive the 
${\rm PA} = \varphi_0+\pi/2$ according to the maximum of the amplitudes $A=A(r)$ in fits $A\sin\left(\varphi-\varphi_0\right)+b$ over all $0\le \varphi \le 2\pi$ for given $r\ge0$, where $b$ refers to an ignorable velocity offset. This procedure is repeated twice with an initial guess of $\rm PA$ based on photometric images in the NSA catalogue. We confirmed that our results are rather insensitive to uncertainty in inclination angles $i$ following various injection experiments.
As shown in Fig. 1, we use a 12\,$^\circ$ width of a PA in scans over $\varphi$ over increments of 1\,$^\circ$. 
Results are found to be robust against choice of this angular width.

In using various photometric parameters including bulge parameters, we choose to use the MaNGA {\em PyMorph Photometric Value Added Catalog} (MPP VAC). 
To probe PA consistency, we downselect samples successfully allowing for a one-component Sersic-fit (in MPP VAC). 
These samples can be presented with one PA extending over the entire galaxy radius, while others are failed to have one-component Seric-fit partly because of bulges and/or bars structures, or other unknown reasons.

This downselection leaves $N=4067$ spiral galaxies.
Focused furthermore on the MaNGA {\em Principal Component Analysis Value Added Catalog} (PCA VAC), leaves a total of $N=4053$ spiral galaxies.

\begin{figure*}
\center{
\includegraphics[scale=0.55]{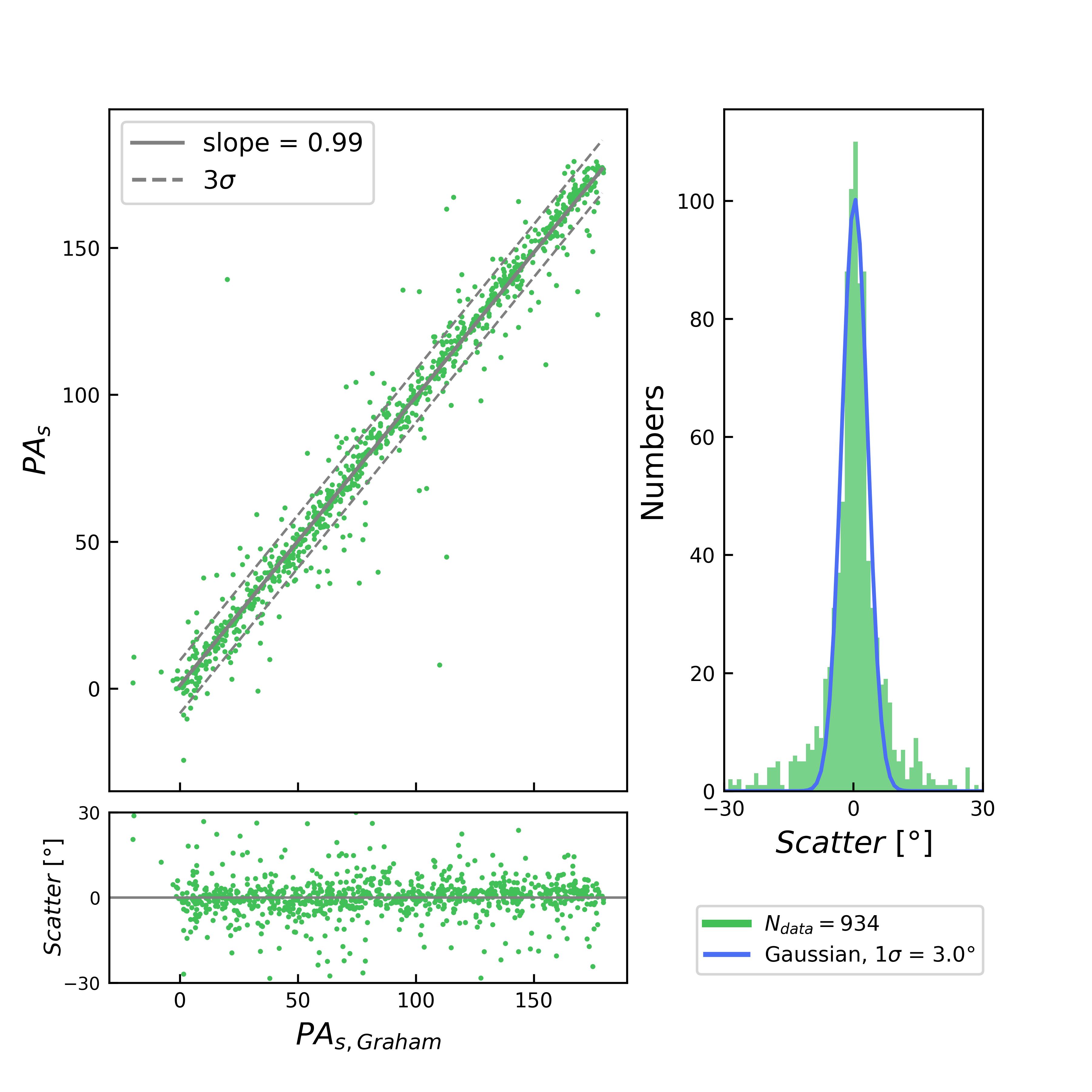}
\includegraphics[scale=0.55]{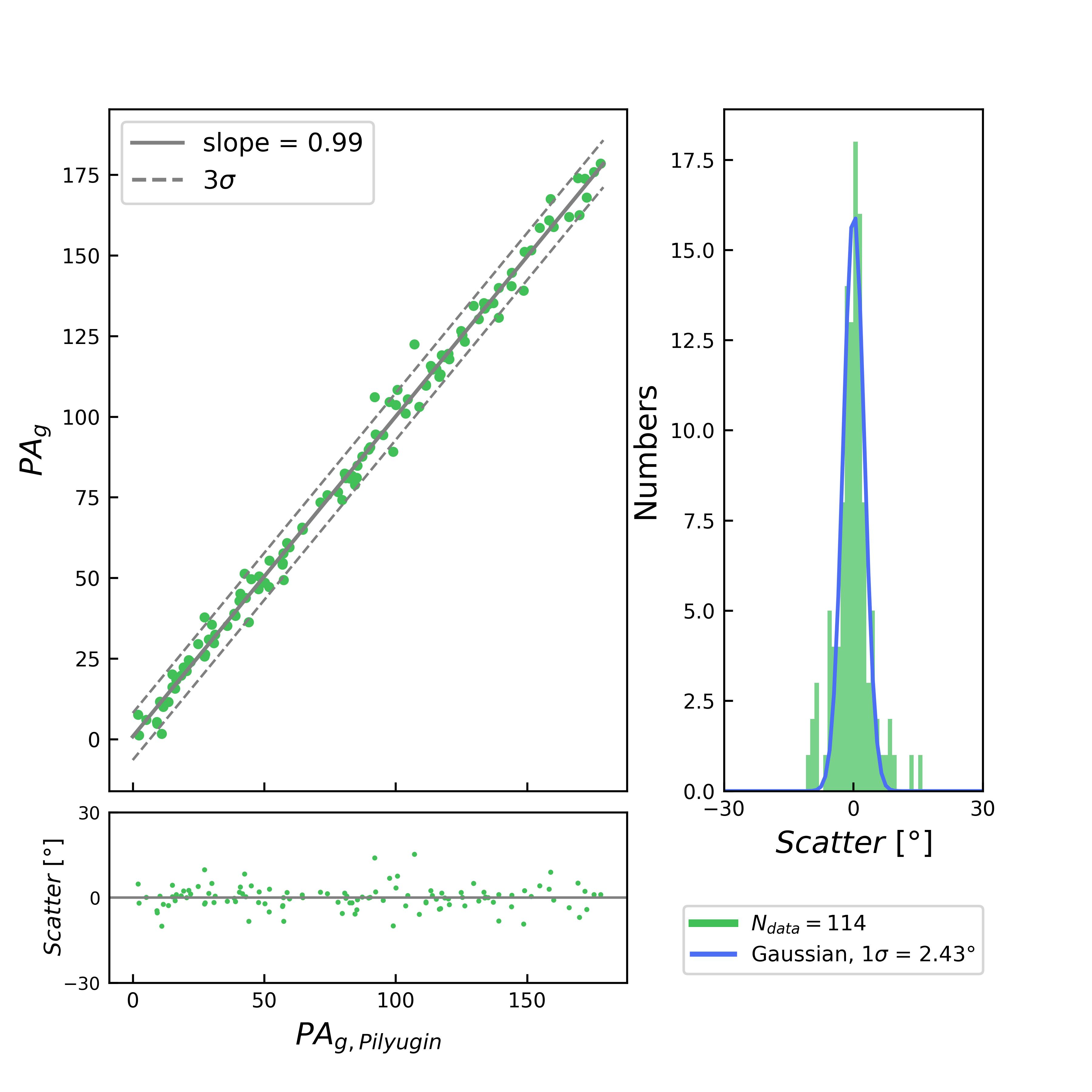}
}
\caption{
(Left panels.) Cross-check of our ${\rm PA}_s$ with ${\rm PA}_{s, Graham}$ from \cite{2018MNRAS.477.4711G}, for 934 galaxies in our sample (Table 1) in common with those of a MaNGA subset of 2721 galaxies in \citep{2018MNRAS.477.4711G}. (Right panels.) Cross-check of our ${\rm PA}_g$ with ${\rm PA}_{g,Pilyugin}$ from \cite{2019A&A...623A.122P}, for 114 galaxies in our sample(Table1) in common with among the sample of \cite{2019A&A...623A.122P}.}
\label{fig:2}
\end{figure*}

As a cross-check, Fig. \ref{fig:2} shows two comparisons of our ${\rm PA}$s with PAs from the literature. 114 galaxies are in common with samples from \cite{2019A&A...623A.122P} which provides 147 MaNGA galaxy properties including  PAs obtained by gas velocity fields. From \cite{2018MNRAS.477.4711G} providing PAs obtained from stellar velocity fields, we cross-checked 934 galaxy PAs of stellar velocity fields.

Fig. 3 shows the resulting 2D distributions of $\Delta{\rm PA}$ obtained from velocity fields (horizontal axis) and photometry (vertical axis) following (\ref{EQN_D1}) and, respectively, 
\begin{eqnarray}
\Delta_2 {\rm PA} = {\rm PA}_{ph} - {\rm PA}_{vf}  
\label{EQN_D2}
\end{eqnarray}
between ${\rm PA}_{ph}$ of photometry and the mean
\begin{eqnarray}
{\rm PA}_{vf}=\frac{1}{2}\left({\rm PA}_g + {\rm PA}_s\right).
\label{EQN_pavf}
\end{eqnarray}
Fig. 3 is restricted to consistency within $30^\circ$, containing 
a remainder of $N=3101$ galaxies detailed in Column 9 of Table 1.

\begin{figure*}
	\center{\includegraphics[scale=0.7]{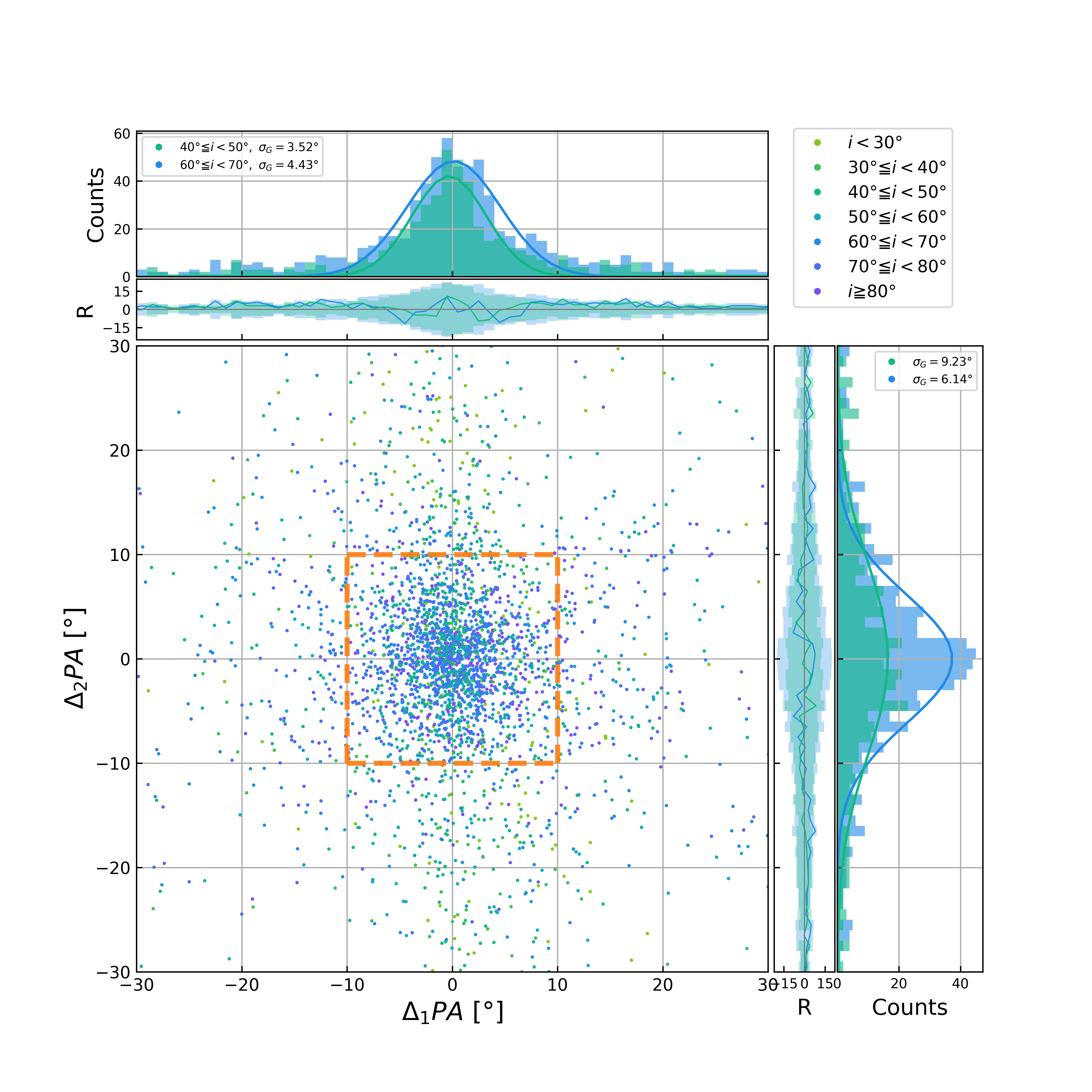}}
    \caption{
    The 2D-distribution of $\Delta_1 {\rm PA}$ in (\ref{EQN_D1}) in the velocity fields of gas and stars (horizontal axis) and $\Delta_2 {\rm PA}$ in (\ref{EQN_D2}) between the PA in photometry and the mean of the two PAs in velocity fields (vertical axis). 
    The distribution is largely Gaussian within $\pm30^\circ$ except for some excess in tails.
    The top and rightmost plot shows the number distributions versus the control parameter $\theta$ according to (\ref{EQN_D1}) and (\ref{EQN_D2}), 
    further showing the accompanying Gaussian fits and $3\sigma$ uncertainties by scatter in Poisson noise.
The dashed orange square indicates a choice of box of $|\Delta_{b} {\rm PA|} < \theta$ (\ref{EQN_b}), 
shown for $\theta=10^\circ$, bounding both 
$\Delta_1{\rm PA}$ and $\Delta_2{\rm PA}$.
}
\label{fig:3}
\end{figure*}

The two traces $\Delta_k{\rm PA}$, $k=1, 2$, of the 2D distribution of $\Delta \rm PA$ can be suitably fitted by Gaussian distributions (side-panels in Fig. 3). The standard deviations $\sigma_G$ of these Gaussian fits tend to be smaller than the standard deviations of the data (Table 2). This indicates that the tails of our distributions are non-Gaussian. It is further noted that this discrepancy in standard deviation appears mostly at relatively small inclination angles and more so in $\Delta_1 \rm PA$ than in $\Delta_2 \rm PA$, even though the latter shows more noise than the former.

We are at liberty to select sub-samples from the 2D distribution of Fig. 3 according to the strips
\begin{eqnarray}
  \left|\Delta_k {\rm PA}\right| < \theta
  \label{EQN_k}
\end{eqnarray}
over $0<\theta<30^\circ$, where $k=1,2$. 
In considering (\ref{EQN_k}) for $\Delta_1{\rm PA}$, $\Delta_{2}{\rm PA}$ is left unbounded and vice versa.
Table 2 lists the sizes $N_i$ and associated standard deviations, derived from the 2D distribution of Fig. 3, further differentiated over a range of inclination angles $i$.

\begin{table*}[ht]
\centering
    \caption{Standard deviations in the 2D distribution of PA angle differences $\Delta_1{\rm PA}$, $\Delta_2{\rm PA}$, and $\Delta_{b}{\rm PA}$ of Fig. 3, listed by bin of inclination $i$. }
    \label{tab:sigma}
\resizebox{\textwidth}{!}{%
    \begin{tabular}{c||c|c|c||c|c||c|c||c|c|c}
    \hline
    & \multicolumn{3}{c||}{All} & \multicolumn{2}{c||}{$|\Delta_1 \mathrm{PA}| < 30^\circ$} & \multicolumn{2}{c||}{$|\Delta_2 \mathrm{PA}| < 30^\circ$} & \multicolumn{3}{c}{$|\Delta_{b} \mathrm{PA}| < 30^\circ$} \\
    \hline
    $i$ & $N_i$ & $\sigma_{(i,\Delta_1{\rm PA})}$ & $\sigma_{(i,\Delta_2{\rm PA})}$ & $N_i$ & $\sigma_{(i,\Delta_1{\rm PA})}$ & $N_i$ & $\sigma_{(i,\Delta_2{\rm PA})}$ & $N_i$ & $\sigma_{(i,\Delta_1{\rm PA})}$ & $\sigma_{(i,\Delta_2{\rm PA})}$ \\
    \hline
    \hline
    $i < 30^\circ$ & 337& 20.1$^\circ$& 42.85$^\circ$& 301 (89\%)& 9.64$^\circ$& 187(55\%)& 15.24$^\circ$& 172 (51\%)& 8.67$^\circ$& 14.68$^\circ$\\
    $30^\circ \leq i < 40^\circ$ & {392} & 21.1$^\circ$& 35.46$^\circ$& 341(87\%)& 9.04$^\circ$& 255(65\%)& 13.01$^\circ$& 237(60\%)& 7.25$^\circ$& 12.67$^\circ$\\
    $40^\circ \leq i < 50^\circ$ & 559& 19.67$^\circ$& 32.48$^\circ$& 506(91\%)& 9.57$^\circ$& 414(74\%)& 11.65$^\circ$& 393(70\%)& 8.56$^\circ$& 11.34$^\circ$\\
    $50^\circ \leq i < 60^\circ$ & 674& 17.53$^\circ$& 27.61$^\circ$& 621(92\%)& 8.94$^\circ$& 550(82\%)& 11.62$^\circ$& 524(78\%)& 8.27$^\circ$& 11.42$^\circ$\\ 
    $60^\circ \leq i < 70^\circ$ & 761& 20.23$^\circ$& 23.39$^\circ$& 685(99\%)& 9.65$^\circ$& 666(88\%)& 9.79$^\circ$& {635(83\%)}& 9.17$^\circ$& 9.02$^\circ$\\
    $70^\circ \leq i < 80^\circ$ &   753& 20.01$^\circ$& 23.09$^\circ$& 679(98\%)& 8.99$^\circ$& 672(89\%)& 8.14$^\circ$& 636(84\%)& 8.31$^\circ$& 7.03$^\circ$\\
    $i \geq 80^\circ$ & 577& 17.62$^\circ$& 19.67$^\circ$& 532
(96\%)& 7.79$^\circ$& 526(91\%)& 7.52$^\circ$& 505(88\%)& 7.38$^\circ$& 6.78$^\circ$ \\
    \hline
    \end{tabular} 
}    
\end{table*}
\begin{figure*}
\center{
\includegraphics[scale=0.55]{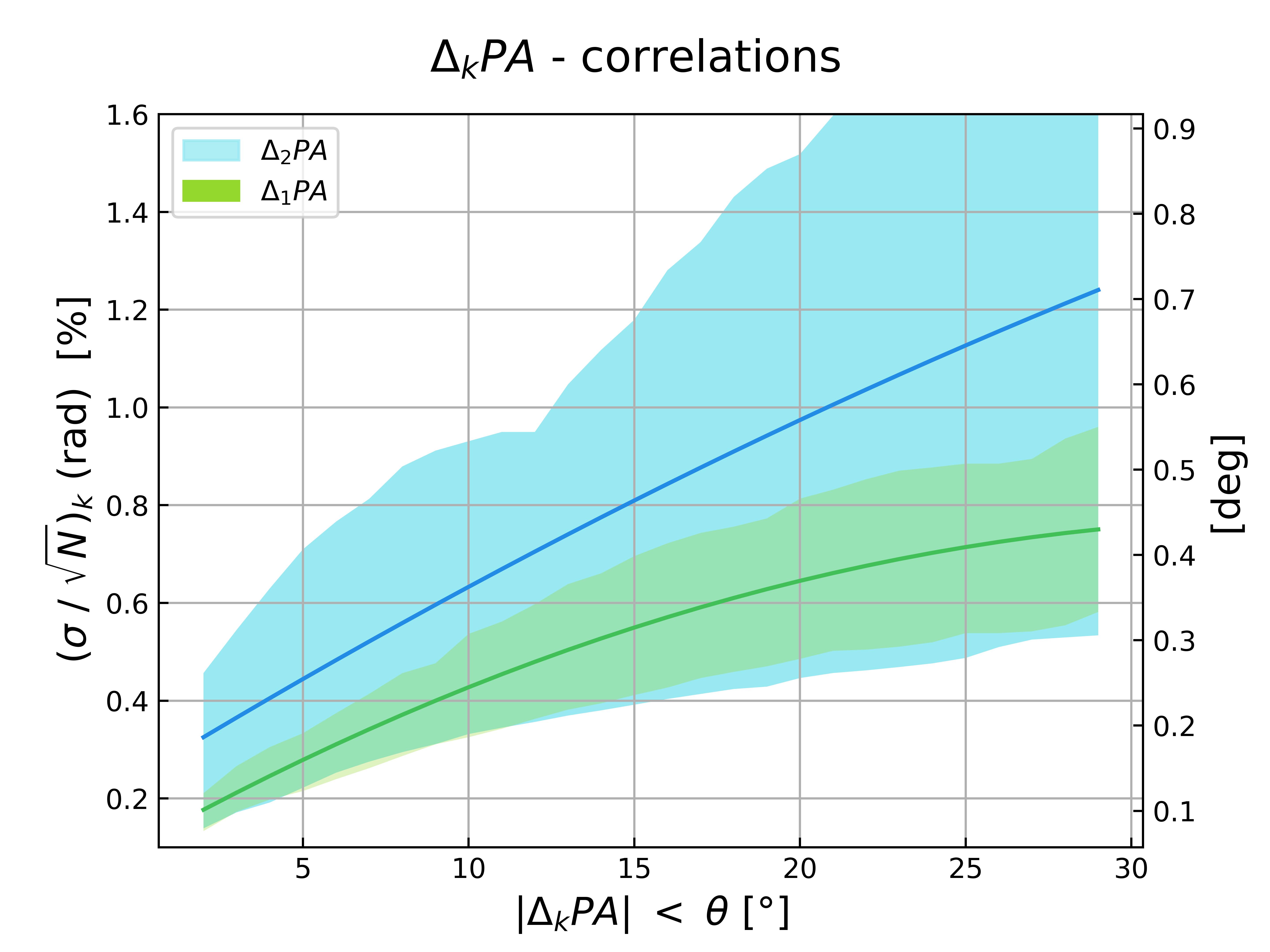} 
\includegraphics[scale=0.55]{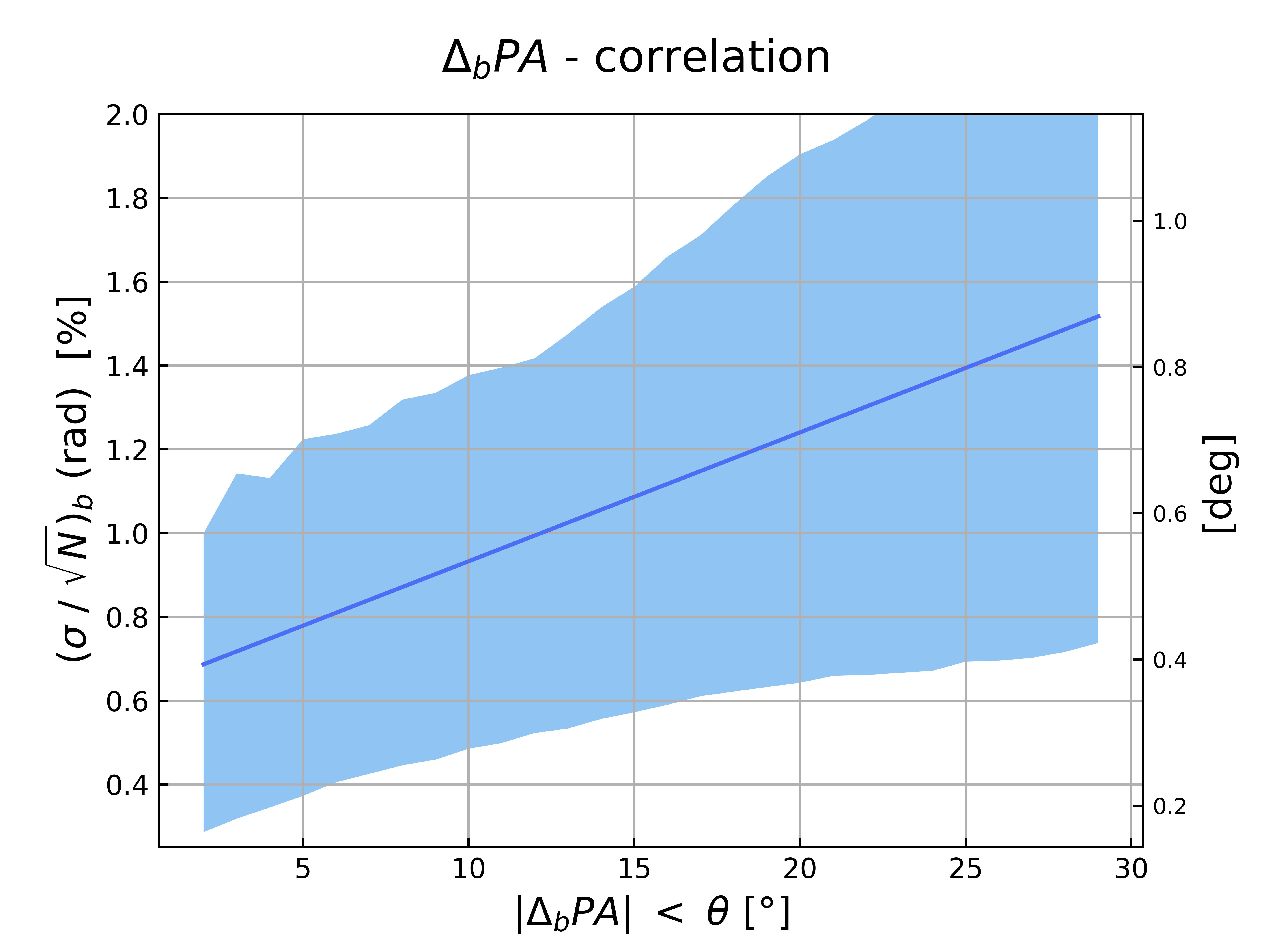} 
}
\caption{Overview of scatter in-the-mean $\sigma/\sqrt{N}$ of the three PAs in MaNGA over $i<80^\circ$ based on Table 2, 
excluding anomalous results for $i\ge80^\circ$. Scatter is defined according to the differences
$\left|\Delta_{k}{\rm PA}\right|<\theta$ $(k=1,2)$ among the three PAs of the photometric image and velocity fields . The black solid lines are fitted line in quadratic formula for left panel, and in linear for right panel. 
The fit coefficients are {\bf provided in (\ref{EQN_fits})}. 
(Left panel.) Scatter as a function of the control parameter $\theta$. While small, PAs of the velocity fields of gas and stars are noticeably more consistent than the same relative to the PA derived from photometry. 
(Right panel.) Scatter within the two-dimensional box $\left|\Delta_{b}{\rm PA}\right|<\theta$ restricting both $\Delta_1{\rm PA}$ and $\Delta_2{\rm PA}$ simultaneously.
The results evidence consistency in MaNGRA sub-samples measured by PAs down to about one percent, providing a suitable starting point for accurate probes of ensemble-averaged rotation curves. The contribution to the total noise budget of $\sigma/\sqrt{N}$
in (\ref{EQN_S}) derives from $\kappa$, determined by varying the control parameter $\theta$.
}
\end{figure*}

The large-$N$ MaNGA sample allows detailed control of $\sigma/\sqrt{N}$ upon selecting sub-samples over strips (\ref{EQN_k}) from the 2D distribution of Fig. 3. For a given ensemble-averaged observable, contributions to a standard error in-the-mean (\ref{EQN_S}) can hereby be explicitly evaluated according to 
\begin{eqnarray}
\kappa_k = \frac{\partial S_E}{\partial \left(\sigma/\sqrt{N}\right)_k},
\label{EQN_K}
\end{eqnarray}
where $k=1,2$.
To quantify this, Fig. 4 shows the scatter in-the-mean, ${\sigma}/{\sqrt{N}}$ with respect to our control parameter $\theta$ in (\ref{EQN_k}). 
The black curves in Fig. 4 represent our results 
in the mean over $i<80^\circ$ with
quadratic fits over $0<\theta<30^\circ$ satisfying
\begin{eqnarray}
\begin{array}{cccc}
\left(\frac{\sigma}{\sqrt{N}}\right)_1 & \simeq & 0.104 + 0.038\, \theta - 5.3\times10^{-4}\theta^2,\\
\left(\frac{\sigma}{\sqrt{N}}\right)_2 & \simeq & 0.244 + 0.041\,\theta -2.4\times10^{-4}\theta^2.
\end{array}
\label{EQN_fits}
\end{eqnarray}

We next consider sub-samples within the box defined by the intersection of the two strips (\ref{EQN_k}). Within this box,
\begin{eqnarray}
    \left| \Delta_b {\rm PA}\right| < \theta,
\label{EQN_b}
\end{eqnarray}
scatter is bounded by $\theta$ simultaneously for both $\Delta_1{\rm PA}$ and $\Delta_2{\rm PA}$.
Table 2 shows $\Delta_1 {\rm PA}$ and $\Delta_2 {\rm PA}$ to be essentially uncorrelated by their small Pearson coefficients. 
Accordingly, we have $\sigma_b \simeq \sqrt{\sigma_1^2+\sigma_2^2}$.
A linear fit to the scatter in the mean over (\ref{EQN_b}) over $\theta < 30^\circ$ (Fig. 4) satisfies
\begin{eqnarray}
\begin{array}{ccc}
\left(\frac{\sigma}{\sqrt{N}}\right)_{b} & \simeq & (0.625 + 0.031 \,\theta^\circ)[\%{\rm rad}].
\end{array}
\label{EQN_box}
\end{eqnarray}
This relation shows control of scatter in PA differences by choice of $0<\theta\le 30^\circ$. In particular, scatter is a mere $1.6\,\%\,{\rm rad}$ at $\theta = 30^\circ$, reducing to $0.94\,\%\,{\rm rad}$ at $\theta = 10^\circ$. 
These values set one-percent lower bounds on a total standard error in-the-mean $S_E$ in ensemble averages. 
In practice, $S_E$ will be larger upon including additional data beyond and independent of the present PAs.

Summarized in Fig. 4 and (\ref{EQN_box}), 
our results on scatter in three PAs demonstrates the potential of MaNGA for high-resolution studies in ensemble-averaged observables of galaxy dynamics on par with modern precision cosmology.

\begin{table*}[ht]
\begin{center}
    \caption{Pearson coefficients $r_{12}$ of $\Delta_1{\rm PA}$ and $\Delta_2{\rm PA}$ of Fig. 3, listed by bin of inclination angle $i$. 
    Results indicate statistical independence of $\Delta_1{\rm PA}$ and $\Delta_2{\rm PA}$ within the expected statistical uncertainty $1/\sqrt{N_i}$ of the Pearson coefficient of size $N_i$ drawn from normal distributions.
    The last three columns refer to the Pearson coefficients $r_{12}$ of $\Delta_1 {\rm PA}$ and $\Delta_2 {\rm PA}$ within the strips (\ref{EQN_k}) of maximal width $\theta = 30^\circ$.
    For the corresponding box of maximal size, correlations remain small on average, though with finite correlations slightly away from zero (last column).
    }
    \vskip0.1in
    \label{tab:incl}
    \begin{tabular}{c||ccc||ccc}
\hline
Inclination $i$& $N_i$ & $r_{12}$ & $r_{12}\sqrt{N_i}$ & $N_i(|\Delta_{b}PA| <30^\circ)$& $r_{12}(|\Delta_{b}PA| <30^\circ)$&$r_{12}\sqrt{N_i}(|\Delta_{b}PA| <30^\circ)$\\
    \hline
    $i < 30^\circ$ 
    & 337& -0.058& -1.0647& 172& 0.1971&2.5849\\
    $30^\circ \leq i < 40^\circ$ & 392&  0.0306&  0.6058& 237& 0.0791&1.2177\\
    $40^\circ \leq i < 50^\circ$ & 559& 0.0291& 0.688& 393& 0.0682&1.352\\
    $50^\circ \leq i < 60^\circ$ & 674& -0.003& --0.0779& 523& 0.112&0.4528\\ 
    $60^\circ \leq i < 70^\circ$ & 761& 0.0758& 2.091& 635& 0.1137&2.8223\\
    $70^\circ \leq i < 80^\circ$ & 753& 0.0314& 0.8616& 636& -0.0106&-0.2673\\
    $i \geq 80^\circ$              & 577& 0.1128& 2.7095& 505& -0.0041&-0.0921\\
    \hline
    \end{tabular}
    \end{center}
\end{table*}

\section{Outlook}

Probing sensitivity of galaxy dynamics at a level of (statistical) uncertainty on par with modern cosmological parameters is a major objective to advancing our understanding of weak gravitation, potentially across all redshifts upon including recent results of JWST.
While $\Lambda$CDM assumes Newtonian gravitation to be scale-free and valid down to arbitrarily small accelerations, conceivably weak gravity in RCs and gravitational collapse producing the first galaxies at cosmic dawn is subject to the background de Sitter scale $a_{dS}$. If so, this offers a prospect of unifying the bTFR and the JWST 'Impossible galaxies' \citep{van24}.

Fig. 4 summarizes our main finding: MaNGA provides the required large $N$ to control scatter in-the-mean of three PAs of photometric images and spectroscopic velocity fields down to the one-percent level. As a minimum to the total uncertainty budget (\ref{EQN_S}), this offers a prospect for high-resolution analysis of stacked MaNGA RCs, provided that the present scatter in PAs provides an effective proxy for any additional sources of noise and uncertainty. 

 In particular, we seek to control scatter in-the-mean in ensemble-averaged rotation curves to accurately resolve the transition to accelerations below $a_{dS}$. In coordinate space, this occurs at the transition radius {\color{blue}\citep{van17a}}
 \begin{eqnarray}
     r_t = \sqrt{R_GR_H}\simeq 4.7\,{\rm kpc}\,M_{11}^{1/2},
     \label{EQN_rt}
 \end{eqnarray}
 where $R_G=GM_b/c^2$ is the gravitational radius of a galaxy of baryonic mass $M_b=M_{11}10^{11}M_\odot$ and $R_H=c/H$ is the Hubble radius, given a Hubble parameter $H=H_0$ at the present epoch. For the Milky Way and similar galaxies, therefore, most of the disk is at accelerations $\alpha < a_{dS}$.
 To confirm previous results, we aim for improved resolution and rigor with MaNGA as an independent and much larger catalog than SPARC.
Stacking of normalized RCs is envisioned in the plane of $a_N/\alpha$ versus $a_N/a_{dS}$, where $a_N$ is the expected Newtonian acceleration based on Newton’s law of gravitation and baryonic mass content and $\alpha=V_c^2/r$ is the centripetal acceleration derived from the presented photometric-spectroscopic data.

By virtue of the large-$N$ MaNGA survey and statistical independence of ${\rm PA}$s (Table 3), contributions of $\Delta_1{\rm PA}$ and $\Delta_2{\rm PA}$ in (\ref{EQN_S}) can be individually identified by their respective $\kappa$ values (\ref{EQN_K}). 
We can also consider their joint contribution, 
considering sub-samples in the box $\left|\Delta_{b}{\rm PA}\right|<\theta$, 
bounding scatter jointly in the $\Delta_1{\rm PA}$ and $\Delta_2{\rm PA}$ with the same $\theta$.
The resulting scatter in-the-mean is about one percent with modest dependence on $\theta$ (Fig. 4). 

Conceivably, scatter in the PAs can be reduced by galaxy modeling, e.g.,
\cite{2020A&A...634A..26P,war73,beg89,deb08,2018MNRAS.473.3256O,2019A&A...623A.122P, 2018MNRAS.473.3256O}. 
Fig. 4, however, shows our present estimates already to satisfy our one percent criterion.

While scatter in PAs is not the only source of noise, we expect it to be a significant contributor to the total noise budget of any observable of ensemble-averaged rotation curves in (\ref{EQN_S}).
Crucially, large sub-samples of MaNGA can be selected within a bound $\theta$ $(\left| \theta \right| < 30^\circ$) (Table 2). 
Their sizes $N_i(\theta)$ remain sufficiently large to permit a detailed measurement of the coefficient $\kappa$ in (\ref{EQN_S}) for each bin of inclination angles $i$ at uncertainties on the order of a few percent by varying the control parameter $\theta$.

Accordingly, MaNGA promises just the kind of large-$N$ survey permitting a high-resolution probe of sensitivity to background cosmology. 
Our focal point is the transition
(\ref{EQN_rt}) of Newtonian to anomalous gravitation in galaxy dynamics across $a_{dS}$, marked by what appears to be a $C^0$-transition in normalized galaxy RCs. At the intended one-percent level of scatter in-the-mean, resolving $a_{dS}$ and hence $H_0$ may be used to distinguish between {\em Planck} $\Lambda$CDM and LDL in light of the $H_0$-tension of about 9\% between the two \citep{van21,rie22}. 
Importantly, $a_{dS}$ and the much smaller scale $a_0$ are related \citep{van24b}
\begin{eqnarray}
    a_0 = \frac{c^2}{2\pi}\sqrt{J}. 
    \label{EQN_a0}
\end{eqnarray}
Here, $J=\left(1-q\right)H^2/c^2$ is the trace of the Schouten tensor \citep{van24,van25} and $q$ denotes the deceleration parameter of the background cosmology. This relationship elucidates redshift dependence in $a_0$ specific to the cosmological background. Indeed, with $H_0$ at hand, (\ref{EQN_a0}) can be used to estimate $q_0$, even though, at present, with very large uncertainty \citep{van24}. 

The present results suggest that MaNGA can provide a confirmation of $a_{dS}$ in the transition to anomalous galaxy dynamics, independent of SPARC, which asymptotes to the bTFR parameterized by $a_0$.
According to our results, stacked MaNGA RCs appear promising to accurately resolve these two distinct scales of acceleration $a_{dS}$ and $a_0$ governing weak gravitation in galaxy dynamics, further across a small but finite range of cosmological redshift.

\mbox{}\\
{\bf CRediT authorship contribution statement}\\

{\bf G.-M. Lee:} Writing, Investigation, Visualization, Formal analysis, Data curation, Validation, Software. 
{\bf M.H.P.M. van Putten:} Conceptualization, Writing — original draft, review \& editing, Validation, Funding acquisition, Supervision.

\mbox{}\\
{\bf Declaration of competing interest}\\

The authors declare that they have no known competing financial
interests or personal relationships that could have appeared to influence the work reported in this paper.

\mbox{}\\
{\bf Acknowledgments} \\

The authors thank the anomous reviewer for a detailed reading and constructive comments.
This work was supported, in part, by NRF grant No. RS-2024-00334550.

\mbox{}\\
{\bf Data Availability}\\

The data underlying this article are available in the MaNGA archives through the MaNGA
online data base.
  


\bibliographystyle{elsarticle-harv} 
\bibliography{main} 

\begin{thebibliography}{56}
\expandafter\ifx\csname natexlab\endcsname\relax\def\natexlab#1{#1}\fi
\providecommand{\url}[1]{\texttt{#1}}
\providecommand{\href}[2]{#2}
\providecommand{\path}[1]{#1}
\providecommand{\DOIprefix}{doi:}
\providecommand{\ArXivprefix}{arXiv:}
\providecommand{\URLprefix}{URL: }
\providecommand{\Pubmedprefix}{pmid:}
\providecommand{\doi}[1]{\href{http://dx.doi.org/#1}{\path{#1}}}
\providecommand{\Pubmed}[1]{\href{pmid:#1}{\path{#1}}}
\providecommand{\bibinfo}[2]{#2}
\ifx\xfnm\relax \def\xfnm[#1]{\unskip,\space#1}\fi
\bibitem[{{Begeman}(1989)}]{beg89}
\bibinfo{author}{{Begeman}, K.G.}, \bibinfo{year}{1989}.
\newblock \bibinfo{title}{{HI rotation curves of spiral galaxies. I. NGC 3198.}}
\newblock \bibinfo{journal}{\aap} \bibinfo{volume}{223}, \bibinfo{pages}{47--60}.
\bibitem[{{Bell} and {de Jong}(2001)}]{bel01}
\bibinfo{author}{{Bell}, E.F.}, \bibinfo{author}{{de Jong}, R.S.}, \bibinfo{year}{2001}.
\newblock \bibinfo{title}{{Stellar Mass-to-Light Ratios and the Tully-Fisher Relation}}.
\newblock \bibinfo{journal}{\apj} \bibinfo{volume}{550}, \bibinfo{pages}{212--229}.
\newblock \DOIprefix\doi{10.1086/319728}, \href{http://arxiv.org/abs/astro-ph/0011493}{{\tt arXiv:astro-ph/0011493}}.
\bibitem[{{Blanton} et~al.(2017){Blanton}, {Bershady}, {Abolfathi}, {Albareti}, {Allende Prieto}, {Almeida}, {Alonso-Garc{\'\i}a}, {Anders}, {Anderson}, {Andrews}, {Aquino-Ort{\'\i}z}, {Arag{\'o}n-Salamanca}, {Argudo-Fern{\'a}ndez}, {Armengaud}, {Aubourg}, {Avila-Reese}, {Badenes}, {Bailey}, {Barger}, {Barrera-Ballesteros}, {Bartosz}, {Bates}, {Baumgarten}, {Bautista}, {Beaton}, {Beers}, {Belfiore}, {Bender}, {Berlind}, {Bernardi}, {Beutler}, {Bird}, {Bizyaev}, {Blanc}, {Blomqvist}, {Bolton}, {Boquien}, {Borissova}, {van den Bosch}, {Bovy}, {Brandt}, {Brinkmann}, {Brownstein}, {Bundy}, {Burgasser}, {Burtin}, {Busca}, {Cappellari}, {Delgado Carigi}, {Carlberg}, {Carnero Rosell}, {Carrera}, {Chanover}, {Cherinka}, {Cheung}, {G{\'o}mez Maqueo Chew}, {Chiappini}, {Choi}, {Chojnowski}, {Chuang}, {Chung}, {Cirolini}, {Clerc}, {Cohen}, {Comparat}, {da Costa}, {Cousinou}, {Covey}, {Crane}, {Croft}, {Cruz-Gonzalez}, {Garrido Cuadra}, {Cunha}, {Damke}, {Darling}, {Davies}, {Dawson}, {de la Macorra}, {Dell'Agli}, {De
  Lee}, {Delubac}, {Di Mille}, {Diamond-Stanic}, {Cano-D{\'\i}az}, {Donor}, {Downes}, {Drory}, {du Mas des Bourboux}, {Duckworth}, {Dwelly}, {Dyer}, {Ebelke}, {Eigenbrot}, {Eisenstein}, {Emsellem}, {Eracleous}, {Escoffier}, {Evans}, {Fan}, {Fern{\'a}ndez-Alvar}, {Fernandez-Trincado}, {Feuillet}, {Finoguenov}, {Fleming}, {Font-Ribera}, {Fredrickson}, {Freischlad}, {Frinchaboy}, {Fuentes}, {Galbany}, {Garcia-Dias}, {Garc{\'\i}a-Hern{\'a}ndez}, {Gaulme}, {Geisler}, {Gelfand}, {Gil-Mar{\'\i}n}, {Gillespie}, {Goddard}, {Gonzalez-Perez}, {Grabowski}, {Green}, {Grier}, {Gunn}, {Guo}, {Guy}, {Hagen}, {Hahn}, {Hall}, {Harding}, {Hasselquist}, {Hawley}, {Hearty}, {Gonzalez Hern{\'a}ndez}, {Ho}, {Hogg}, {Holley-Bockelmann}, {Holtzman}, {Holzer}, {Huehnerhoff}, {Hutchinson}, {Hwang}, {Ibarra-Medel}, {da Silva Ilha}, {Ivans}, {Ivory}, {Jackson}, {Jensen}, {Johnson}, {Jones}, {J{\"o}nsson}, {Jullo}, {Kamble}, {Kinemuchi}, {Kirkby}, {Kitaura}, {Klaene}, {Knapp}, {Kneib}, {Kollmeier}, {Lacerna}, {Lane}, {Lang}, {Law},
  {Lazarz}, {Lee}, {Le Goff}, {Liang}, {Li}, {Li}, {Lian}, {Lima}, {Lin}, {Lin}, {Bertran de Lis}, {Liu}, {de Icaza Lizaola}, {Long}, {Lucatello}, {Lundgren}, {MacDonald}, {Deconto Machado}, {MacLeod}, {Mahadevan}, {Geimba Maia}, {Maiolino}, {Majewski}, {Malanushenko}, {Malanushenko}, {Manchado}, {Mao}, {Maraston}, {Marques-Chaves}, {Masseron}, {Masters}, {McBride}, {McDermid}, {McGrath}, {McGreer}, {Medina Pe{\~n}a}, {Melendez}, {Merloni}, {Merrifield}, {Meszaros}, {Meza}, {Minchev}, {Minniti}, {Miyaji}, {More}, {Mulchaey}, {M{\"u}ller-S{\'a}nchez}, {Muna}, {Munoz}, {Myers}, {Nair}, {Nandra}, {Correa do Nascimento}, {Negrete}, {Ness}, {Newman}, {Nichol}, {Nidever}, {Nitschelm}, {Ntelis}, {O'Connell}, {Oelkers}, {Oravetz}, {Oravetz}, {Pace}, {Padilla}, {Palanque-Delabrouille}, {Alonso Palicio}, {Pan}, {Parejko}, {Parikh}, {P{\^a}ris}, {Park}, {Patten}, {Peirani}, {Pellejero-Ibanez}, {Penny}, {Percival}, {Perez-Fournon}, {Petitjean}, {Pieri}, {Pinsonneault}, {Pisani}, {Poleski}, {Prada}, {Prakash}, {Queiroz},
  {Raddick}, {Raichoor}, {Barboza Rembold}, {Richstein}, {Riffel}, {Riffel}, {Rix}, {Robin}, {Rockosi}, {Rodr{\'\i}guez-Torres}, {Roman-Lopes}, {Rom{\'a}n-Z{\'u}{\~n}iga}, {Rosado}, {Ross}, {Rossi}, {Ruan}, {Ruggeri}, {Rykoff}, {Salazar-Albornoz}, {Salvato}, {S{\'a}nchez}, {Aguado}, {S{\'a}nchez-Gallego}, {Santana}, {Santiago}, {Sayres}, {Schiavon}, {da Silva Schimoia}, {Schlafly}, {Schlegel}, {Schneider}, {Schultheis}, {Schuster}, {Schwope}, {Seo}, {Shao}, {Shen}, {Shetrone}, {Shull}, {Simon}, {Skinner}, {Skrutskie}, {Slosar}, {Smith}, {Sobeck}, {Sobreira}, {Somers}, {Souto}, {Stark}, {Stassun}, {Stauffer}, {Steinmetz}, {Storchi-Bergmann}, {Streblyanska}, {Stringfellow}, {Su{\'a}rez}, {Sun}, {Suzuki}, {Szigeti}, {Taghizadeh-Popp}, {Tang}, {Tao}, {Tayar}, {Tembe}, {Teske}, {Thakar}, {Thomas}, {Thompson}, {Tinker}, {Tissera}, {Tojeiro}, {Hernandez Toledo}, {de la Torre}, {Tremonti}, {Troup}, {Valenzuela}, {Martinez Valpuesta}, {Vargas-Gonz{\'a}lez}, {Vargas-Maga{\~n}a}, {Vazquez}, {Villanova}, {Vivek}, {Vogt},
  {Wake}, {Walterbos}, {Wang}, {Weaver}, {Weijmans}, {Weinberg}, {Westfall}, {Whelan}, {Wild}, {Wilson}, {Wood-Vasey}, {Wylezalek}, {Xiao}, {Yan}, {Yang}, {Ybarra}, {Y{\`e}che}, {Zakamska}, {Zamora}, {Zarrouk}, {Zasowski}, {Zhang}, {Zhao}, {Zheng}, {Zheng}, {Zhou}, {Zhou}, {Zhu}, {Zoccali} and {Zou}}]{2017AJ....154...28B}
\bibinfo{author}{{Blanton}, M.R.}, \bibinfo{author}{{Bershady}, M.A.}, \bibinfo{author}{{Abolfathi}, B.}, \bibinfo{author}{{Albareti}, F.D.}, \bibinfo{author}{{Allende Prieto}, C.}, \bibinfo{author}{{Almeida}, A.}, \bibinfo{author}{{Alonso-Garc{\'\i}a}, J.}, \bibinfo{author}{{Anders}, F.}, \bibinfo{author}{{Anderson}, S.F.}, \bibinfo{author}{{Andrews}, B.}, \bibinfo{author}{{Aquino-Ort{\'\i}z}, E.}, \bibinfo{author}{{Arag{\'o}n-Salamanca}, A.}, \bibinfo{author}{{Argudo-Fern{\'a}ndez}, M.}, \bibinfo{author}{{Armengaud}, E.}, \bibinfo{author}{{Aubourg}, E.}, \bibinfo{author}{{Avila-Reese}, V.}, \bibinfo{author}{{Badenes}, C.}, \bibinfo{author}{{Bailey}, S.}, \bibinfo{author}{{Barger}, K.A.}, \bibinfo{author}{{Barrera-Ballesteros}, J.}, \bibinfo{author}{{Bartosz}, C.}, \bibinfo{author}{{Bates}, D.}, \bibinfo{author}{{Baumgarten}, F.}, \bibinfo{author}{{Bautista}, J.}, \bibinfo{author}{{Beaton}, R.}, \bibinfo{author}{{Beers}, T.C.}, \bibinfo{author}{{Belfiore}, F.}, \bibinfo{author}{{Bender}, C.F.},
  \bibinfo{author}{{Berlind}, A.A.}, \bibinfo{author}{{Bernardi}, M.}, \bibinfo{author}{{Beutler}, F.}, \bibinfo{author}{{Bird}, J.C.}, \bibinfo{author}{{Bizyaev}, D.}, \bibinfo{author}{{Blanc}, G.A.}, \bibinfo{author}{{Blomqvist}, M.}, \bibinfo{author}{{Bolton}, A.S.}, \bibinfo{author}{{Boquien}, M.}, \bibinfo{author}{{Borissova}, J.}, \bibinfo{author}{{van den Bosch}, R.}, \bibinfo{author}{{Bovy}, J.}, \bibinfo{author}{{Brandt}, W.N.}, \bibinfo{author}{{Brinkmann}, J.}, \bibinfo{author}{{Brownstein}, J.R.}, \bibinfo{author}{{Bundy}, K.}, \bibinfo{author}{{Burgasser}, A.J.}, \bibinfo{author}{{Burtin}, E.}, \bibinfo{author}{{Busca}, N.G.}, \bibinfo{author}{{Cappellari}, M.}, \bibinfo{author}{{Delgado Carigi}, M.L.}, \bibinfo{author}{{Carlberg}, J.K.}, \bibinfo{author}{{Carnero Rosell}, A.}, \bibinfo{author}{{Carrera}, R.}, \bibinfo{author}{{Chanover}, N.J.}, \bibinfo{author}{{Cherinka}, B.}, \bibinfo{author}{{Cheung}, E.}, \bibinfo{author}{{G{\'o}mez Maqueo Chew}, Y.}, \bibinfo{author}{{Chiappini}, C.},
  \bibinfo{author}{{Choi}, P.D.}, \bibinfo{author}{{Chojnowski}, D.}, \bibinfo{author}{{Chuang}, C.H.}, \bibinfo{author}{{Chung}, H.}, \bibinfo{author}{{Cirolini}, R.F.}, \bibinfo{author}{{Clerc}, N.}, \bibinfo{author}{{Cohen}, R.E.}, \bibinfo{author}{{Comparat}, J.}, \bibinfo{author}{{da Costa}, L.}, \bibinfo{author}{{Cousinou}, M.C.}, \bibinfo{author}{{Covey}, K.}, \bibinfo{author}{{Crane}, J.D.}, \bibinfo{author}{{Croft}, R.A.C.}, \bibinfo{author}{{Cruz-Gonzalez}, I.}, \bibinfo{author}{{Garrido Cuadra}, D.}, \bibinfo{author}{{Cunha}, K.}, \bibinfo{author}{{Damke}, G.J.}, \bibinfo{author}{{Darling}, J.}, \bibinfo{author}{{Davies}, R.}, \bibinfo{author}{{Dawson}, K.}, \bibinfo{author}{{de la Macorra}, A.}, \bibinfo{author}{{Dell'Agli}, F.}, \bibinfo{author}{{De Lee}, N.}, \bibinfo{author}{{Delubac}, T.}, \bibinfo{author}{{Di Mille}, F.}, \bibinfo{author}{{Diamond-Stanic}, A.}, \bibinfo{author}{{Cano-D{\'\i}az}, M.}, \bibinfo{author}{{Donor}, J.}, \bibinfo{author}{{Downes}, J.J.}, \bibinfo{author}{{Drory},
  N.}, \bibinfo{author}{{du Mas des Bourboux}, H.}, \bibinfo{author}{{Duckworth}, C.J.}, \bibinfo{author}{{Dwelly}, T.}, \bibinfo{author}{{Dyer}, J.}, \bibinfo{author}{{Ebelke}, G.}, \bibinfo{author}{{Eigenbrot}, A.D.}, \bibinfo{author}{{Eisenstein}, D.J.}, \bibinfo{author}{{Emsellem}, E.}, \bibinfo{author}{{Eracleous}, M.}, \bibinfo{author}{{Escoffier}, S.}, \bibinfo{author}{{Evans}, M.L.}, \bibinfo{author}{{Fan}, X.}, \bibinfo{author}{{Fern{\'a}ndez-Alvar}, E.}, \bibinfo{author}{{Fernandez-Trincado}, J.G.}, \bibinfo{author}{{Feuillet}, D.K.}, \bibinfo{author}{{Finoguenov}, A.}, \bibinfo{author}{{Fleming}, S.W.}, \bibinfo{author}{{Font-Ribera}, A.}, \bibinfo{author}{{Fredrickson}, A.}, \bibinfo{author}{{Freischlad}, G.}, \bibinfo{author}{{Frinchaboy}, P.M.}, \bibinfo{author}{{Fuentes}, C.E.}, \bibinfo{author}{{Galbany}, L.}, \bibinfo{author}{{Garcia-Dias}, R.}, \bibinfo{author}{{Garc{\'\i}a-Hern{\'a}ndez}, D.A.}, \bibinfo{author}{{Gaulme}, P.}, \bibinfo{author}{{Geisler}, D.}, \bibinfo{author}{{Gelfand},
  J.D.}, \bibinfo{author}{{Gil-Mar{\'\i}n}, H.}, \bibinfo{author}{{Gillespie}, B.A.}, \bibinfo{author}{{Goddard}, D.}, \bibinfo{author}{{Gonzalez-Perez}, V.}, \bibinfo{author}{{Grabowski}, K.}, \bibinfo{author}{{Green}, P.J.}, \bibinfo{author}{{Grier}, C.J.}, \bibinfo{author}{{Gunn}, J.E.}, \bibinfo{author}{{Guo}, H.}, \bibinfo{author}{{Guy}, J.}, \bibinfo{author}{{Hagen}, A.}, \bibinfo{author}{{Hahn}, C.}, \bibinfo{author}{{Hall}, M.}, \bibinfo{author}{{Harding}, P.}, \bibinfo{author}{{Hasselquist}, S.}, \bibinfo{author}{{Hawley}, S.L.}, \bibinfo{author}{{Hearty}, F.}, \bibinfo{author}{{Gonzalez Hern{\'a}ndez}, J.I.}, \bibinfo{author}{{Ho}, S.}, \bibinfo{author}{{Hogg}, D.W.}, \bibinfo{author}{{Holley-Bockelmann}, K.}, \bibinfo{author}{{Holtzman}, J.A.}, \bibinfo{author}{{Holzer}, P.H.}, \bibinfo{author}{{Huehnerhoff}, J.}, \bibinfo{author}{{Hutchinson}, T.A.}, \bibinfo{author}{{Hwang}, H.S.}, \bibinfo{author}{{Ibarra-Medel}, H.J.}, \bibinfo{author}{{da Silva Ilha}, G.}, \bibinfo{author}{{Ivans}, I.I.},
  \bibinfo{author}{{Ivory}, K.}, \bibinfo{author}{{Jackson}, K.}, \bibinfo{author}{{Jensen}, T.W.}, \bibinfo{author}{{Johnson}, J.A.}, \bibinfo{author}{{Jones}, A.}, \bibinfo{author}{{J{\"o}nsson}, H.}, \bibinfo{author}{{Jullo}, E.}, \bibinfo{author}{{Kamble}, V.}, \bibinfo{author}{{Kinemuchi}, K.}, \bibinfo{author}{{Kirkby}, D.}, \bibinfo{author}{{Kitaura}, F.S.}, \bibinfo{author}{{Klaene}, M.}, \bibinfo{author}{{Knapp}, G.R.}, \bibinfo{author}{{Kneib}, J.P.}, \bibinfo{author}{{Kollmeier}, J.A.}, \bibinfo{author}{{Lacerna}, I.}, \bibinfo{author}{{Lane}, R.R.}, \bibinfo{author}{{Lang}, D.}, \bibinfo{author}{{Law}, D.R.}, \bibinfo{author}{{Lazarz}, D.}, \bibinfo{author}{{Lee}, Y.}, \bibinfo{author}{{Le Goff}, J.M.}, \bibinfo{author}{{Liang}, F.H.}, \bibinfo{author}{{Li}, C.}, \bibinfo{author}{{Li}, H.}, \bibinfo{author}{{Lian}, J.}, \bibinfo{author}{{Lima}, M.}, \bibinfo{author}{{Lin}, L.}, \bibinfo{author}{{Lin}, Y.T.}, \bibinfo{author}{{Bertran de Lis}, S.}, \bibinfo{author}{{Liu}, C.}, \bibinfo{author}{{de
  Icaza Lizaola}, M.A.C.}, \bibinfo{author}{{Long}, D.}, \bibinfo{author}{{Lucatello}, S.}, \bibinfo{author}{{Lundgren}, B.}, \bibinfo{author}{{MacDonald}, N.K.}, \bibinfo{author}{{Deconto Machado}, A.}, \bibinfo{author}{{MacLeod}, C.L.}, \bibinfo{author}{{Mahadevan}, S.}, \bibinfo{author}{{Geimba Maia}, M.A.}, \bibinfo{author}{{Maiolino}, R.}, \bibinfo{author}{{Majewski}, S.R.}, \bibinfo{author}{{Malanushenko}, E.}, \bibinfo{author}{{Malanushenko}, V.}, \bibinfo{author}{{Manchado}, A.}, \bibinfo{author}{{Mao}, S.}, \bibinfo{author}{{Maraston}, C.}, \bibinfo{author}{{Marques-Chaves}, R.}, \bibinfo{author}{{Masseron}, T.}, \bibinfo{author}{{Masters}, K.L.}, \bibinfo{author}{{McBride}, C.K.}, \bibinfo{author}{{McDermid}, R.M.}, \bibinfo{author}{{McGrath}, B.}, \bibinfo{author}{{McGreer}, I.D.}, \bibinfo{author}{{Medina Pe{\~n}a}, N.}, \bibinfo{author}{{Melendez}, M.}, \bibinfo{author}{{Merloni}, A.}, \bibinfo{author}{{Merrifield}, M.R.}, \bibinfo{author}{{Meszaros}, S.}, \bibinfo{author}{{Meza}, A.},
  \bibinfo{author}{{Minchev}, I.}, \bibinfo{author}{{Minniti}, D.}, \bibinfo{author}{{Miyaji}, T.}, \bibinfo{author}{{More}, S.}, \bibinfo{author}{{Mulchaey}, J.}, \bibinfo{author}{{M{\"u}ller-S{\'a}nchez}, F.}, \bibinfo{author}{{Muna}, D.}, \bibinfo{author}{{Munoz}, R.R.}, \bibinfo{author}{{Myers}, A.D.}, \bibinfo{author}{{Nair}, P.}, \bibinfo{author}{{Nandra}, K.}, \bibinfo{author}{{Correa do Nascimento}, J.}, \bibinfo{author}{{Negrete}, A.}, \bibinfo{author}{{Ness}, M.}, \bibinfo{author}{{Newman}, J.A.}, \bibinfo{author}{{Nichol}, R.C.}, \bibinfo{author}{{Nidever}, D.L.}, \bibinfo{author}{{Nitschelm}, C.}, \bibinfo{author}{{Ntelis}, P.}, \bibinfo{author}{{O'Connell}, J.E.}, \bibinfo{author}{{Oelkers}, R.J.}, \bibinfo{author}{{Oravetz}, A.}, \bibinfo{author}{{Oravetz}, D.}, \bibinfo{author}{{Pace}, Z.}, \bibinfo{author}{{Padilla}, N.}, \bibinfo{author}{{Palanque-Delabrouille}, N.}, \bibinfo{author}{{Alonso Palicio}, P.}, \bibinfo{author}{{Pan}, K.}, \bibinfo{author}{{Parejko}, J.K.},
  \bibinfo{author}{{Parikh}, T.}, \bibinfo{author}{{P{\^a}ris}, I.}, \bibinfo{author}{{Park}, C.}, \bibinfo{author}{{Patten}, A.Y.}, \bibinfo{author}{{Peirani}, S.}, \bibinfo{author}{{Pellejero-Ibanez}, M.}, \bibinfo{author}{{Penny}, S.}, \bibinfo{author}{{Percival}, W.J.}, \bibinfo{author}{{Perez-Fournon}, I.}, \bibinfo{author}{{Petitjean}, P.}, \bibinfo{author}{{Pieri}, M.M.}, \bibinfo{author}{{Pinsonneault}, M.}, \bibinfo{author}{{Pisani}, A.}, \bibinfo{author}{{Poleski}, R.}, \bibinfo{author}{{Prada}, F.}, \bibinfo{author}{{Prakash}, A.}, \bibinfo{author}{{Queiroz}, A.B.d.A.}, \bibinfo{author}{{Raddick}, M.J.}, \bibinfo{author}{{Raichoor}, A.}, \bibinfo{author}{{Barboza Rembold}, S.}, \bibinfo{author}{{Richstein}, H.}, \bibinfo{author}{{Riffel}, R.A.}, \bibinfo{author}{{Riffel}, R.}, \bibinfo{author}{{Rix}, H.W.}, \bibinfo{author}{{Robin}, A.C.}, \bibinfo{author}{{Rockosi}, C.M.}, \bibinfo{author}{{Rodr{\'\i}guez-Torres}, S.}, \bibinfo{author}{{Roman-Lopes}, A.},
  \bibinfo{author}{{Rom{\'a}n-Z{\'u}{\~n}iga}, C.}, \bibinfo{author}{{Rosado}, M.}, \bibinfo{author}{{Ross}, A.J.}, \bibinfo{author}{{Rossi}, G.}, \bibinfo{author}{{Ruan}, J.}, \bibinfo{author}{{Ruggeri}, R.}, \bibinfo{author}{{Rykoff}, E.S.}, \bibinfo{author}{{Salazar-Albornoz}, S.}, \bibinfo{author}{{Salvato}, M.}, \bibinfo{author}{{S{\'a}nchez}, A.G.}, \bibinfo{author}{{Aguado}, D.S.}, \bibinfo{author}{{S{\'a}nchez-Gallego}, J.R.}, \bibinfo{author}{{Santana}, F.A.}, \bibinfo{author}{{Santiago}, B.X.}, \bibinfo{author}{{Sayres}, C.}, \bibinfo{author}{{Schiavon}, R.P.}, \bibinfo{author}{{da Silva Schimoia}, J.}, \bibinfo{author}{{Schlafly}, E.F.}, \bibinfo{author}{{Schlegel}, D.J.}, \bibinfo{author}{{Schneider}, D.P.}, \bibinfo{author}{{Schultheis}, M.}, \bibinfo{author}{{Schuster}, W.J.}, \bibinfo{author}{{Schwope}, A.}, \bibinfo{author}{{Seo}, H.J.}, \bibinfo{author}{{Shao}, Z.}, \bibinfo{author}{{Shen}, S.}, \bibinfo{author}{{Shetrone}, M.}, \bibinfo{author}{{Shull}, M.}, \bibinfo{author}{{Simon}, J.D.},
  \bibinfo{author}{{Skinner}, D.}, \bibinfo{author}{{Skrutskie}, M.F.}, \bibinfo{author}{{Slosar}, A.}, \bibinfo{author}{{Smith}, V.V.}, \bibinfo{author}{{Sobeck}, J.S.}, \bibinfo{author}{{Sobreira}, F.}, \bibinfo{author}{{Somers}, G.}, \bibinfo{author}{{Souto}, D.}, \bibinfo{author}{{Stark}, D.V.}, \bibinfo{author}{{Stassun}, K.}, \bibinfo{author}{{Stauffer}, F.}, \bibinfo{author}{{Steinmetz}, M.}, \bibinfo{author}{{Storchi-Bergmann}, T.}, \bibinfo{author}{{Streblyanska}, A.}, \bibinfo{author}{{Stringfellow}, G.S.}, \bibinfo{author}{{Su{\'a}rez}, G.}, \bibinfo{author}{{Sun}, J.}, \bibinfo{author}{{Suzuki}, N.}, \bibinfo{author}{{Szigeti}, L.}, \bibinfo{author}{{Taghizadeh-Popp}, M.}, \bibinfo{author}{{Tang}, B.}, \bibinfo{author}{{Tao}, C.}, \bibinfo{author}{{Tayar}, J.}, \bibinfo{author}{{Tembe}, M.}, \bibinfo{author}{{Teske}, J.}, \bibinfo{author}{{Thakar}, A.R.}, \bibinfo{author}{{Thomas}, D.}, \bibinfo{author}{{Thompson}, B.A.}, \bibinfo{author}{{Tinker}, J.L.}, \bibinfo{author}{{Tissera}, P.},
  \bibinfo{author}{{Tojeiro}, R.}, \bibinfo{author}{{Hernandez Toledo}, H.}, \bibinfo{author}{{de la Torre}, S.}, \bibinfo{author}{{Tremonti}, C.}, \bibinfo{author}{{Troup}, N.W.}, \bibinfo{author}{{Valenzuela}, O.}, \bibinfo{author}{{Martinez Valpuesta}, I.}, \bibinfo{author}{{Vargas-Gonz{\'a}lez}, J.}, \bibinfo{author}{{Vargas-Maga{\~n}a}, M.}, \bibinfo{author}{{Vazquez}, J.A.}, \bibinfo{author}{{Villanova}, S.}, \bibinfo{author}{{Vivek}, M.}, \bibinfo{author}{{Vogt}, N.}, \bibinfo{author}{{Wake}, D.}, \bibinfo{author}{{Walterbos}, R.}, \bibinfo{author}{{Wang}, Y.}, \bibinfo{author}{{Weaver}, B.A.}, \bibinfo{author}{{Weijmans}, A.M.}, \bibinfo{author}{{Weinberg}, D.H.}, \bibinfo{author}{{Westfall}, K.B.}, \bibinfo{author}{{Whelan}, D.G.}, \bibinfo{author}{{Wild}, V.}, \bibinfo{author}{{Wilson}, J.}, \bibinfo{author}{{Wood-Vasey}, W.M.}, \bibinfo{author}{{Wylezalek}, D.}, \bibinfo{author}{{Xiao}, T.}, \bibinfo{author}{{Yan}, R.}, \bibinfo{author}{{Yang}, M.}, \bibinfo{author}{{Ybarra}, J.E.},
  \bibinfo{author}{{Y{\`e}che}, C.}, \bibinfo{author}{{Zakamska}, N.}, \bibinfo{author}{{Zamora}, O.}, \bibinfo{author}{{Zarrouk}, P.}, \bibinfo{author}{{Zasowski}, G.}, \bibinfo{author}{{Zhang}, K.}, \bibinfo{author}{{Zhao}, G.B.}, \bibinfo{author}{{Zheng}, Z.}, \bibinfo{author}{{Zheng}, Z.}, \bibinfo{author}{{Zhou}, X.}, \bibinfo{author}{{Zhou}, Z.M.}, \bibinfo{author}{{Zhu}, G.B.}, \bibinfo{author}{{Zoccali}, M.}, \bibinfo{author}{{Zou}, H.}, \bibinfo{year}{2017}.
\newblock \bibinfo{title}{{Sloan Digital Sky Survey IV: Mapping the Milky Way, Nearby Galaxies, and the Distant Universe}}.
\newblock \bibinfo{journal}{\aj} \bibinfo{volume}{154}, \bibinfo{pages}{28}.
\newblock \DOIprefix\doi{10.3847/1538-3881/aa7567}, \href{http://arxiv.org/abs/1703.00052}{{\tt arXiv:1703.00052}}.
\bibitem[{{Brinchmann} et~al.(2004){Brinchmann}, {Charlot}, {White}, {Tremonti}, {Kauffmann}, {Heckman} and {Brinkmann}}]{bri04}
\bibinfo{author}{{Brinchmann}, J.}, \bibinfo{author}{{Charlot}, S.}, \bibinfo{author}{{White}, S.D.M.}, \bibinfo{author}{{Tremonti}, C.}, \bibinfo{author}{{Kauffmann}, G.}, \bibinfo{author}{{Heckman}, T.}, \bibinfo{author}{{Brinkmann}, J.}, \bibinfo{year}{2004}.
\newblock \bibinfo{title}{{The physical properties of star-forming galaxies in the low-redshift Universe}}.
\newblock \bibinfo{journal}{\mnras} \bibinfo{volume}{351}, \bibinfo{pages}{1151--1179}.
\newblock \DOIprefix\doi{10.1111/j.1365-2966.2004.07881.x}, \href{http://arxiv.org/abs/astro-ph/0311060}{{\tt arXiv:astro-ph/0311060}}.
\bibitem[{{Bundy} et~al.(2015){Bundy}, {Bershady}, {Law}, {Yan}, {Drory}, {MacDonald}, {Wake}, {Cherinka}, {S{\'a}nchez-Gallego}, {Weijmans}, {Thomas}, {Tremonti}, {Masters}, {Coccato}, {Diamond-Stanic}, {Arag{\'o}n-Salamanca}, {Avila-Reese}, {Badenes}, {Falc{\'o}n-Barroso}, {Belfiore}, {Bizyaev}, {Blanc}, {Bland-Hawthorn}, {Blanton}, {Brownstein}, {Byler}, {Cappellari}, {Conroy}, {Dutton}, {Emsellem}, {Etherington}, {Frinchaboy}, {Fu}, {Gunn}, {Harding}, {Johnston}, {Kauffmann}, {Kinemuchi}, {Klaene}, {Knapen}, {Leauthaud}, {Li}, {Lin}, {Maiolino}, {Malanushenko}, {Malanushenko}, {Mao}, {Maraston}, {McDermid}, {Merrifield}, {Nichol}, {Oravetz}, {Pan}, {Parejko}, {Sanchez}, {Schlegel}, {Simmons}, {Steele}, {Steinmetz}, {Thanjavur}, {Thompson}, {Tinker}, {van den Bosch}, {Westfall}, {Wilkinson}, {Wright}, {Xiao} and {Zhang}}]{2015ApJ...798....7B}
\bibinfo{author}{{Bundy}, K.}, \bibinfo{author}{{Bershady}, M.A.}, \bibinfo{author}{{Law}, D.R.}, \bibinfo{author}{{Yan}, R.}, \bibinfo{author}{{Drory}, N.}, \bibinfo{author}{{MacDonald}, N.}, \bibinfo{author}{{Wake}, D.A.}, \bibinfo{author}{{Cherinka}, B.}, \bibinfo{author}{{S{\'a}nchez-Gallego}, J.R.}, \bibinfo{author}{{Weijmans}, A.M.}, \bibinfo{author}{{Thomas}, D.}, \bibinfo{author}{{Tremonti}, C.}, \bibinfo{author}{{Masters}, K.}, \bibinfo{author}{{Coccato}, L.}, \bibinfo{author}{{Diamond-Stanic}, A.M.}, \bibinfo{author}{{Arag{\'o}n-Salamanca}, A.}, \bibinfo{author}{{Avila-Reese}, V.}, \bibinfo{author}{{Badenes}, C.}, \bibinfo{author}{{Falc{\'o}n-Barroso}, J.}, \bibinfo{author}{{Belfiore}, F.}, \bibinfo{author}{{Bizyaev}, D.}, \bibinfo{author}{{Blanc}, G.A.}, \bibinfo{author}{{Bland-Hawthorn}, J.}, \bibinfo{author}{{Blanton}, M.R.}, \bibinfo{author}{{Brownstein}, J.R.}, \bibinfo{author}{{Byler}, N.}, \bibinfo{author}{{Cappellari}, M.}, \bibinfo{author}{{Conroy}, C.}, \bibinfo{author}{{Dutton}, A.A.},
  \bibinfo{author}{{Emsellem}, E.}, \bibinfo{author}{{Etherington}, J.}, \bibinfo{author}{{Frinchaboy}, P.M.}, \bibinfo{author}{{Fu}, H.}, \bibinfo{author}{{Gunn}, J.E.}, \bibinfo{author}{{Harding}, P.}, \bibinfo{author}{{Johnston}, E.J.}, \bibinfo{author}{{Kauffmann}, G.}, \bibinfo{author}{{Kinemuchi}, K.}, \bibinfo{author}{{Klaene}, M.A.}, \bibinfo{author}{{Knapen}, J.H.}, \bibinfo{author}{{Leauthaud}, A.}, \bibinfo{author}{{Li}, C.}, \bibinfo{author}{{Lin}, L.}, \bibinfo{author}{{Maiolino}, R.}, \bibinfo{author}{{Malanushenko}, V.}, \bibinfo{author}{{Malanushenko}, E.}, \bibinfo{author}{{Mao}, S.}, \bibinfo{author}{{Maraston}, C.}, \bibinfo{author}{{McDermid}, R.M.}, \bibinfo{author}{{Merrifield}, M.R.}, \bibinfo{author}{{Nichol}, R.C.}, \bibinfo{author}{{Oravetz}, D.}, \bibinfo{author}{{Pan}, K.}, \bibinfo{author}{{Parejko}, J.K.}, \bibinfo{author}{{Sanchez}, S.F.}, \bibinfo{author}{{Schlegel}, D.}, \bibinfo{author}{{Simmons}, A.}, \bibinfo{author}{{Steele}, O.}, \bibinfo{author}{{Steinmetz}, M.},
  \bibinfo{author}{{Thanjavur}, K.}, \bibinfo{author}{{Thompson}, B.A.}, \bibinfo{author}{{Tinker}, J.L.}, \bibinfo{author}{{van den Bosch}, R.C.E.}, \bibinfo{author}{{Westfall}, K.B.}, \bibinfo{author}{{Wilkinson}, D.}, \bibinfo{author}{{Wright}, S.}, \bibinfo{author}{{Xiao}, T.}, \bibinfo{author}{{Zhang}, K.}, \bibinfo{year}{2015}.
\newblock \bibinfo{title}{{Overview of the SDSS-IV MaNGA Survey: Mapping nearby Galaxies at Apache Point Observatory}}.
\newblock \bibinfo{journal}{\apj} \bibinfo{volume}{798}, \bibinfo{pages}{7}.
\newblock \DOIprefix\doi{10.1088/0004-637X/798/1/7}, \href{http://arxiv.org/abs/1412.1482}{{\tt arXiv:1412.1482}}.
\bibitem[{{Cappellari} and {Copin}(2003)}]{2003MNRAS.342..345C}
\bibinfo{author}{{Cappellari}, M.}, \bibinfo{author}{{Copin}, Y.}, \bibinfo{year}{2003}.
\newblock \bibinfo{title}{{Adaptive spatial binning of integral-field spectroscopic data using Voronoi tessellations}}.
\newblock \bibinfo{journal}{\mnras} \bibinfo{volume}{342}, \bibinfo{pages}{345--354}.
\newblock \DOIprefix\doi{10.1046/j.1365-8711.2003.06541.x}, \href{http://arxiv.org/abs/astro-ph/0302262}{{\tt arXiv:astro-ph/0302262}}.
\bibitem[{{Conroy}(2013)}]{Conroy13}
\bibinfo{author}{{Conroy}, C.}, \bibinfo{year}{2013}.
\newblock \bibinfo{title}{{Modeling the Panchromatic Spectral Energy Distributions of Galaxies}}.
\newblock \bibinfo{journal}{\araa} \bibinfo{volume}{51}, \bibinfo{pages}{393--455}.
\newblock \DOIprefix\doi{10.1146/annurev-astro-082812-141017}, \href{http://arxiv.org/abs/1301.7095}{{\tt arXiv:1301.7095}}.
\bibitem[{{de Blok} et~al.(2008){de Blok}, {Walter}, {Brinks}, {Trachternach}, {Oh} and {Kennicutt}}]{deb08}
\bibinfo{author}{{de Blok}, W.J.G.}, \bibinfo{author}{{Walter}, F.}, \bibinfo{author}{{Brinks}, E.}, \bibinfo{author}{{Trachternach}, C.}, \bibinfo{author}{{Oh}, S.H.}, \bibinfo{author}{{Kennicutt}, R.~C., J.}, \bibinfo{year}{2008}.
\newblock \bibinfo{title}{{High-Resolution Rotation Curves and Galaxy Mass Models from THINGS}}.
\newblock \bibinfo{journal}{\aj} \bibinfo{volume}{136}, \bibinfo{pages}{2648--2719}.
\newblock \DOIprefix\doi{10.1088/0004-6256/136/6/2648}, \href{http://arxiv.org/abs/0810.2100}{{\tt arXiv:0810.2100}}.
\bibitem[{{Di Valentino} et~al.(2021){Di Valentino}, {Mena}, {Pan}, {Visinelli}, {Yang}, {Melchiorri}, {Mota}, {Riess} and {Silk}}]{div21}
\bibinfo{author}{{Di Valentino}, E.}, \bibinfo{author}{{Mena}, O.}, \bibinfo{author}{{Pan}, S.}, \bibinfo{author}{{Visinelli}, L.}, \bibinfo{author}{{Yang}, W.}, \bibinfo{author}{{Melchiorri}, A.}, \bibinfo{author}{{Mota}, D.F.}, \bibinfo{author}{{Riess}, A.G.}, \bibinfo{author}{{Silk}, J.}, \bibinfo{year}{2021}.
\newblock \bibinfo{title}{{In the realm of the Hubble tension-a review of solutions}}.
\newblock \bibinfo{journal}{Classical and Quantum Gravity} \bibinfo{volume}{38}, \bibinfo{pages}{153001}.
\newblock \DOIprefix\doi{10.1088/1361-6382/ac086d}, \href{http://arxiv.org/abs/2103.01183}{{\tt arXiv:2103.01183}}.
\bibitem[{{Dom{\'\i}nguez S{\'a}nchez} et~al.(2018){Dom{\'\i}nguez S{\'a}nchez}, {Huertas-Company}, {Bernardi}, {Tuccillo} and {Fischer}}]{2018MNRAS.476.3661D}
\bibinfo{author}{{Dom{\'\i}nguez S{\'a}nchez}, H.}, \bibinfo{author}{{Huertas-Company}, M.}, \bibinfo{author}{{Bernardi}, M.}, \bibinfo{author}{{Tuccillo}, D.}, \bibinfo{author}{{Fischer}, J.L.}, \bibinfo{year}{2018}.
\newblock \bibinfo{title}{{Improving galaxy morphologies for SDSS with Deep Learning}}.
\newblock \bibinfo{journal}{\mnras} \bibinfo{volume}{476}, \bibinfo{pages}{3661--3676}.
\newblock \DOIprefix\doi{10.1093/mnras/sty338}, \href{http://arxiv.org/abs/1711.05744}{{\tt arXiv:1711.05744}}.
\bibitem[{{Dom{\'\i}nguez S{\'a}nchez} et~al.(2022){Dom{\'\i}nguez S{\'a}nchez}, {Margalef}, {Bernardi} and {Huertas-Company}}]{2022MNRAS.509.4024D}
\bibinfo{author}{{Dom{\'\i}nguez S{\'a}nchez}, H.}, \bibinfo{author}{{Margalef}, B.}, \bibinfo{author}{{Bernardi}, M.}, \bibinfo{author}{{Huertas-Company}, M.}, \bibinfo{year}{2022}.
\newblock \bibinfo{title}{{SDSS-IV DR17: final release of MaNGA PyMorph photometric and deep-learning morphological catalogues}}.
\newblock \bibinfo{journal}{\mnras} \bibinfo{volume}{509}, \bibinfo{pages}{4024--4036}.
\newblock \DOIprefix\doi{10.1093/mnras/stab3089}, \href{http://arxiv.org/abs/2110.10694}{{\tt arXiv:2110.10694}}.
\bibitem[{{Famaey} and {McGaugh}(2012)}]{fam12}
\bibinfo{author}{{Famaey}, B.}, \bibinfo{author}{{McGaugh}, S.S.}, \bibinfo{year}{2012}.
\newblock \bibinfo{title}{{Modified Newtonian Dynamics (MOND): Observational Phenomenology and Relativistic Extensions}}.
\newblock \bibinfo{journal}{Living Reviews in Relativity} \bibinfo{volume}{15}, \bibinfo{pages}{10}.
\newblock \DOIprefix\doi{10.12942/lrr-2012-10}, \href{http://arxiv.org/abs/1112.3960}{{\tt arXiv:1112.3960}}.
\bibitem[{{Graham} et~al.(2018){Graham}, {Cappellari}, {Li}, {Mao}, {Bershady}, {Bizyaev}, {Brinkmann}, {Brownstein}, {Bundy}, {Drory}, {Law}, {Pan}, {Thomas}, {Wake}, {Weijmans}, {Westfall} and {Yan}}]{2018MNRAS.477.4711G}
\bibinfo{author}{{Graham}, M.T.}, \bibinfo{author}{{Cappellari}, M.}, \bibinfo{author}{{Li}, H.}, \bibinfo{author}{{Mao}, S.}, \bibinfo{author}{{Bershady}, M.A.}, \bibinfo{author}{{Bizyaev}, D.}, \bibinfo{author}{{Brinkmann}, J.}, \bibinfo{author}{{Brownstein}, J.R.}, \bibinfo{author}{{Bundy}, K.}, \bibinfo{author}{{Drory}, N.}, \bibinfo{author}{{Law}, D.R.}, \bibinfo{author}{{Pan}, K.}, \bibinfo{author}{{Thomas}, D.}, \bibinfo{author}{{Wake}, D.A.}, \bibinfo{author}{{Weijmans}, A.M.}, \bibinfo{author}{{Westfall}, K.B.}, \bibinfo{author}{{Yan}, R.}, \bibinfo{year}{2018}.
\newblock \bibinfo{title}{{SDSS-IV MaNGA: stellar angular momentum of about 2300 galaxies: unveiling the bimodality of massive galaxy properties}}.
\newblock \bibinfo{journal}{\mnras} \bibinfo{volume}{477}, \bibinfo{pages}{4711--4737}.
\newblock \DOIprefix\doi{10.1093/mnras/sty504}, \href{http://arxiv.org/abs/1802.08213}{{\tt arXiv:1802.08213}}.
\bibitem[{{Hibat-Allah} et~al.(2023){Hibat-Allah}, {Melko} and {Carrasquilla}}]{hib23}
\bibinfo{author}{{Hibat-Allah}, M.}, \bibinfo{author}{{Melko}, R.G.}, \bibinfo{author}{{Carrasquilla}, J.}, \bibinfo{year}{2023}.
\newblock \bibinfo{title}{{Investigating topological order using recurrent neural networks}}.
\newblock \bibinfo{journal}{\prb} \bibinfo{volume}{108}, \bibinfo{pages}{075152}.
\newblock \DOIprefix\doi{10.1103/PhysRevB.108.075152}, \href{http://arxiv.org/abs/2303.11207}{{\tt arXiv:2303.11207}}.
\bibitem[{{Huang} et~al.(2024){Huang}, {Li} and {Yu}}]{hua24}
\bibinfo{author}{{Huang}, F.}, \bibinfo{author}{{Li}, Y.Z.}, \bibinfo{author}{{Yu}, J.H.}, \bibinfo{year}{2024}.
\newblock \bibinfo{title}{{Distinguishing thermal histories of dark matter from structure formation}}.
\newblock \bibinfo{journal}{\jcap} \bibinfo{volume}{2024}, \bibinfo{pages}{023}.
\newblock \DOIprefix\doi{10.1088/1475-7516/2024/01/023}, \href{http://arxiv.org/abs/2306.00065}{{\tt arXiv:2306.00065}}.
\bibitem[{{Hubble}(1927)}]{1927Obs....50..276H}
\bibinfo{author}{{Hubble}, E.P.}, \bibinfo{year}{1927}.
\newblock \bibinfo{title}{{The classification of spiral nebulae}}.
\newblock \bibinfo{journal}{The Observatory} \bibinfo{volume}{50}, \bibinfo{pages}{276--281}.
\bibitem[{{Jedamzik} et~al.(2021){Jedamzik}, {Pogosian} and {Zhao}}]{2021CmPhy...4..123J}
\bibinfo{author}{{Jedamzik}, K.}, \bibinfo{author}{{Pogosian}, L.}, \bibinfo{author}{{Zhao}, G.B.}, \bibinfo{year}{2021}.
\newblock \bibinfo{title}{{Why reducing the cosmic sound horizon alone can not fully resolve the Hubble tension}}.
\newblock \bibinfo{journal}{Communications Physics} \bibinfo{volume}{4}, \bibinfo{pages}{123}.
\newblock \DOIprefix\doi{10.1038/s42005-021-00628-x}, \href{http://arxiv.org/abs/2010.04158}{{\tt arXiv:2010.04158}}.
\bibitem[{{Ku}(1966)}]{ku66}
\bibinfo{author}{{Ku}, H.H.}, \bibinfo{year}{1966}.
\newblock \bibinfo{title}{{Notes on the use of propagation of error formulas}}.
\newblock \bibinfo{journal}{J. Res. Nat. B. Stand.} \bibinfo{volume}{70C}, \bibinfo{pages}{262}.
\newblock \DOIprefix\doi{10.6028/jres.070c.025}.
\bibitem[{{Lelli} et~al.(2016){Lelli}, {McGaugh} and {Schombert}}]{lel16}
\bibinfo{author}{{Lelli}, F.}, \bibinfo{author}{{McGaugh}, S.S.}, \bibinfo{author}{{Schombert}, J.M.}, \bibinfo{year}{2016}.
\newblock \bibinfo{title}{{SPARC: Mass Models for 175 Disk Galaxies with Spitzer Photometry and Accurate Rotation Curves}}.
\newblock \bibinfo{journal}{\aj} \bibinfo{volume}{152}, \bibinfo{pages}{157}.
\newblock \DOIprefix\doi{10.3847/0004-6256/152/6/157}, \href{http://arxiv.org/abs/1606.09251}{{\tt arXiv:1606.09251}}.
\bibitem[{{Lu} et~al.(2023){Lu}, {Zhu}, {Cappellari}, {Li}, {Mao} and {Xu}}]{lu23}
\bibinfo{author}{{Lu}, S.}, \bibinfo{author}{{Zhu}, K.}, \bibinfo{author}{{Cappellari}, M.}, \bibinfo{author}{{Li}, R.}, \bibinfo{author}{{Mao}, S.}, \bibinfo{author}{{Xu}, D.}, \bibinfo{year}{2023}.
\newblock \bibinfo{title}{{MaNGA DynPop - II. Global stellar population, gradients, and star-formation histories from integral-field spectroscopy of 10K galaxies: link with galaxy rotation, shape, and total-density gradients}}.
\newblock \bibinfo{journal}{\mnras} \bibinfo{volume}{526}, \bibinfo{pages}{1022--1045}.
\newblock \DOIprefix\doi{10.1093/mnras/stad2732}, \href{http://arxiv.org/abs/2304.11712}{{\tt arXiv:2304.11712}}.
\bibitem[{{Lu} et~al.(2024){Lu}, {Zhu}, {Cappellari}, {Li}, {Mao} and {Xu}}]{lu24}
\bibinfo{author}{{Lu}, S.}, \bibinfo{author}{{Zhu}, K.}, \bibinfo{author}{{Cappellari}, M.}, \bibinfo{author}{{Li}, R.}, \bibinfo{author}{{Mao}, S.}, \bibinfo{author}{{Xu}, D.}, \bibinfo{year}{2024}.
\newblock \bibinfo{title}{{MaNGA DynPop - V. The dark-matter fraction versus stellar velocity dispersion relation and stellar initial mass function variations in galaxies: dynamical models and full spectrum fitting of integral-field spectroscopy}}.
\newblock \bibinfo{journal}{\mnras} \bibinfo{volume}{530}, \bibinfo{pages}{4474--4492}.
\newblock \DOIprefix\doi{10.1093/mnras/stae1116}, \href{http://arxiv.org/abs/2309.12395}{{\tt arXiv:2309.12395}}.
\bibitem[{{Madau} and {Dickinson}(2014)}]{mad14}
\bibinfo{author}{{Madau}, P.}, \bibinfo{author}{{Dickinson}, M.}, \bibinfo{year}{2014}.
\newblock \bibinfo{title}{{Cosmic Star-Formation History}}.
\newblock \bibinfo{journal}{\araa} \bibinfo{volume}{52}, \bibinfo{pages}{415--486}.
\newblock \DOIprefix\doi{10.1146/annurev-astro-081811-125615}, \href{http://arxiv.org/abs/1403.0007}{{\tt arXiv:1403.0007}}.
\bibitem[{{Maslej-Kre{\v{s}}{\v{n}}{\'a}kov{\'a}} et~al.(2021){Maslej-Kre{\v{s}}{\v{n}}{\'a}kov{\'a}}, {El Bouchefry} and {Butka}}]{mas21}
\bibinfo{author}{{Maslej-Kre{\v{s}}{\v{n}}{\'a}kov{\'a}}, V.}, \bibinfo{author}{{El Bouchefry}, K.}, \bibinfo{author}{{Butka}, P.}, \bibinfo{year}{2021}.
\newblock \bibinfo{title}{{Morphological classification of compact and extended radio galaxies using convolutional neural networks and data augmentation techniques}}.
\newblock \bibinfo{journal}{\mnras} \bibinfo{volume}{505}, \bibinfo{pages}{1464--1475}.
\newblock \DOIprefix\doi{10.1093/mnras/stab1400}, \href{http://arxiv.org/abs/2107.00385}{{\tt arXiv:2107.00385}}.
\bibitem[{{McGaugh}(2012)}]{mcg12}
\bibinfo{author}{{McGaugh}, S.S.}, \bibinfo{year}{2012}.
\newblock \bibinfo{title}{{The Baryonic Tully-Fisher Relation of Gas-rich Galaxies as a Test of {\ensuremath{\Lambda}}CDM and MOND}}.
\newblock \bibinfo{journal}{\aj} \bibinfo{volume}{143}, \bibinfo{pages}{40}.
\newblock \DOIprefix\doi{10.1088/0004-6256/143/2/40}, \href{http://arxiv.org/abs/1107.2934}{{\tt arXiv:1107.2934}}.
\bibitem[{{McGaugh} et~al.(2016){McGaugh}, {Lelli} and {Schombert}}]{mcg16}
\bibinfo{author}{{McGaugh}, S.S.}, \bibinfo{author}{{Lelli}, F.}, \bibinfo{author}{{Schombert}, J.M.}, \bibinfo{year}{2016}.
\newblock \bibinfo{title}{{Radial Acceleration Relation in Rotationally Supported Galaxies}}.
\newblock \bibinfo{journal}{\prl} \bibinfo{volume}{117}, \bibinfo{pages}{201101}.
\newblock \DOIprefix\doi{10.1103/PhysRevLett.117.201101}, \href{http://arxiv.org/abs/1609.05917}{{\tt arXiv:1609.05917}}.
\bibitem[{{McGaugh} et~al.(2000){McGaugh}, {Schombert}, {Bothun} and {de Blok}}]{mcg00}
\bibinfo{author}{{McGaugh}, S.S.}, \bibinfo{author}{{Schombert}, J.M.}, \bibinfo{author}{{Bothun}, G.D.}, \bibinfo{author}{{de Blok}, W.J.G.}, \bibinfo{year}{2000}.
\newblock \bibinfo{title}{{The Baryonic Tully-Fisher Relation}}.
\newblock \bibinfo{journal}{\apjl} \bibinfo{volume}{533}, \bibinfo{pages}{L99--L102}.
\newblock \DOIprefix\doi{10.1086/312628}, \href{http://arxiv.org/abs/astro-ph/0003001}{{\tt arXiv:astro-ph/0003001}}.
\bibitem[{{Medina-Rosales} et~al.(2024){Medina-Rosales}, {Cabrera-Vives} and {Miller}}]{med24}
\bibinfo{author}{{Medina-Rosales}, E.}, \bibinfo{author}{{Cabrera-Vives}, G.}, \bibinfo{author}{{Miller}, C.J.}, \bibinfo{year}{2024}.
\newblock \bibinfo{title}{{Mitigating bias in deep learning: training unbiased models on biased data for the morphological classification of galaxies}}.
\newblock \bibinfo{journal}{\mnras} \bibinfo{volume}{531}, \bibinfo{pages}{52--60}.
\newblock \DOIprefix\doi{10.1093/mnras/stae1088}, \href{http://arxiv.org/abs/2308.11007}{{\tt arXiv:2308.11007}}.
\bibitem[{{Milgrom}(1983)}]{mil83}
\bibinfo{author}{{Milgrom}, M.}, \bibinfo{year}{1983}.
\newblock \bibinfo{title}{{A modification of the Newtonian dynamics as a possible alternative to the hidden mass hypothesis.}}
\newblock \bibinfo{journal}{\apj} \bibinfo{volume}{270}, \bibinfo{pages}{365--370}.
\newblock \DOIprefix\doi{10.1086/161130}.
\bibitem[{{Natarajan} et~al.(2024){Natarajan}, {Williams}, {Brada{\v{c}}}, {Grillo}, {Ghosh}, {Sharon} and {Wagner}}]{nat24}
\bibinfo{author}{{Natarajan}, P.}, \bibinfo{author}{{Williams}, L.L.R.}, \bibinfo{author}{{Brada{\v{c}}}, M.}, \bibinfo{author}{{Grillo}, C.}, \bibinfo{author}{{Ghosh}, A.}, \bibinfo{author}{{Sharon}, K.}, \bibinfo{author}{{Wagner}, J.}, \bibinfo{year}{2024}.
\newblock \bibinfo{title}{{Strong Lensing by Galaxy Clusters}}.
\newblock \bibinfo{journal}{\ssr} \bibinfo{volume}{220}, \bibinfo{pages}{19}.
\newblock \DOIprefix\doi{10.1007/s11214-024-01051-8}, \href{http://arxiv.org/abs/2403.06245}{{\tt arXiv:2403.06245}}.
\bibitem[{{Ndung'u} et~al.(2023){Ndung'u}, {Grobler}, {Wijnholds}, {Karastoyanova} and {Azzopardi}}]{ndu23}
\bibinfo{author}{{Ndung'u}, S.}, \bibinfo{author}{{Grobler}, T.}, \bibinfo{author}{{Wijnholds}, S.J.}, \bibinfo{author}{{Karastoyanova}, D.}, \bibinfo{author}{{Azzopardi}, G.}, \bibinfo{year}{2023}.
\newblock \bibinfo{title}{{Advances on the morphological classification of radio galaxies: A review}}.
\newblock \bibinfo{journal}{\nar} \bibinfo{volume}{97}, \bibinfo{pages}{101685}.
\newblock \DOIprefix\doi{10.1016/j.newar.2023.101685}.
\bibitem[{{Nestor Shachar} et~al.(2023){Nestor Shachar}, {Price}, {F{\"o}rster Schreiber}, {Genzel}, {Shimizu}, {Tacconi}, {{\"U}bler}, {Burkert}, {Davies}, {Dekel}, {Herrera-Camus}, {Lee}, {Liu}, {Lutz}, {Naab}, {Neri}, {Renzini}, {Saglia}, {Schuster}, {Sternberg}, {Wisnioski} and {Wuyts}}]{nes23}
\bibinfo{author}{{Nestor Shachar}, A.}, \bibinfo{author}{{Price}, S.H.}, \bibinfo{author}{{F{\"o}rster Schreiber}, N.M.}, \bibinfo{author}{{Genzel}, R.}, \bibinfo{author}{{Shimizu}, T.T.}, \bibinfo{author}{{Tacconi}, L.J.}, \bibinfo{author}{{{\"U}bler}, H.}, \bibinfo{author}{{Burkert}, A.}, \bibinfo{author}{{Davies}, R.I.}, \bibinfo{author}{{Dekel}, A.}, \bibinfo{author}{{Herrera-Camus}, R.}, \bibinfo{author}{{Lee}, L.L.}, \bibinfo{author}{{Liu}, D.}, \bibinfo{author}{{Lutz}, D.}, \bibinfo{author}{{Naab}, T.}, \bibinfo{author}{{Neri}, R.}, \bibinfo{author}{{Renzini}, A.}, \bibinfo{author}{{Saglia}, R.}, \bibinfo{author}{{Schuster}, K.F.}, \bibinfo{author}{{Sternberg}, A.}, \bibinfo{author}{{Wisnioski}, E.}, \bibinfo{author}{{Wuyts}, S.}, \bibinfo{year}{2023}.
\newblock \bibinfo{title}{{RC100: Rotation Curves of 100 Massive Star-forming Galaxies at z = 0.6-2.5 Reveal Little Dark Matter on Galactic Scales}}.
\newblock \bibinfo{journal}{\apj} \bibinfo{volume}{944}, \bibinfo{pages}{78}.
\newblock \DOIprefix\doi{10.3847/1538-4357/aca9cf}, \href{http://arxiv.org/abs/2209.12199}{{\tt arXiv:2209.12199}}.
\bibitem[{{Oh} et~al.(2018){Oh}, {Staveley-Smith}, {Spekkens}, {Kamphuis} and {Koribalski}}]{2018MNRAS.473.3256O}
\bibinfo{author}{{Oh}, S.H.}, \bibinfo{author}{{Staveley-Smith}, L.}, \bibinfo{author}{{Spekkens}, K.}, \bibinfo{author}{{Kamphuis}, P.}, \bibinfo{author}{{Koribalski}, B.S.}, \bibinfo{year}{2018}.
\newblock \bibinfo{title}{{2D Bayesian automated tilted-ring fitting of disc galaxies in large H I galaxy surveys: 2DBAT}}.
\newblock \bibinfo{journal}{\mnras} \bibinfo{volume}{473}, \bibinfo{pages}{3256--3298}.
\newblock \DOIprefix\doi{10.1093/mnras/stx2304}, \href{http://arxiv.org/abs/1709.02049}{{\tt arXiv:1709.02049}}.
\bibitem[{{Perivolaropoulos} and {Skara}(2022)}]{per22}
\bibinfo{author}{{Perivolaropoulos}, L.}, \bibinfo{author}{{Skara}, F.}, \bibinfo{year}{2022}.
\newblock \bibinfo{title}{{Challenges for {\ensuremath{\Lambda}}CDM: An update}}.
\newblock \bibinfo{journal}{\nar} \bibinfo{volume}{95}, \bibinfo{pages}{101659}.
\newblock \DOIprefix\doi{10.1016/j.newar.2022.101659}, \href{http://arxiv.org/abs/2105.05208}{{\tt arXiv:2105.05208}}.
\bibitem[{{Pilyugin} et~al.(2019){Pilyugin}, {Grebel}, {Zinchenko}, {Nefedyev} and {V{\'\i}lchez}}]{2019A&A...623A.122P}
\bibinfo{author}{{Pilyugin}, L.S.}, \bibinfo{author}{{Grebel}, E.K.}, \bibinfo{author}{{Zinchenko}, I.A.}, \bibinfo{author}{{Nefedyev}, Y.A.}, \bibinfo{author}{{V{\'\i}lchez}, J.M.}, \bibinfo{year}{2019}.
\newblock \bibinfo{title}{{Relations between abundance characteristics and rotation velocity for star-forming MaNGA galaxies}}.
\newblock \bibinfo{journal}{\aap} \bibinfo{volume}{623}, \bibinfo{pages}{A122}.
\newblock \DOIprefix\doi{10.1051/0004-6361/201834239}, \href{http://arxiv.org/abs/1901.11001}{{\tt arXiv:1901.11001}}.
\bibitem[{{Pilyugin} et~al.(2020){Pilyugin}, {Grebel}, {Zinchenko}, {V{\'\i}lchez}, {Sakhibov}, {Nefedyev} and {Berczik}}]{2020A&A...634A..26P}
\bibinfo{author}{{Pilyugin}, L.S.}, \bibinfo{author}{{Grebel}, E.K.}, \bibinfo{author}{{Zinchenko}, I.A.}, \bibinfo{author}{{V{\'\i}lchez}, J.M.}, \bibinfo{author}{{Sakhibov}, F.}, \bibinfo{author}{{Nefedyev}, Y.A.}, \bibinfo{author}{{Berczik}, P.P.}, \bibinfo{year}{2020}.
\newblock \bibinfo{title}{{Properties of galaxies with an offset between the position angles of the major kinematic and photometric axes}}.
\newblock \bibinfo{journal}{\aap} \bibinfo{volume}{634}, \bibinfo{pages}{A26}.
\newblock \DOIprefix\doi{10.1051/0004-6361/201936357}, \href{http://arxiv.org/abs/1912.03233}{{\tt arXiv:1912.03233}}.
\bibitem[{{Riess} et~al.(2022){Riess}, {Yuan}, {Macri}, {Scolnic}, {Brout}, {Casertano}, {Jones}, {Murakami}, {Anand}, {Breuval}, {Brink}, {Filippenko}, {Hoffmann}, {Jha}, {D'arcy Kenworthy}, {Mackenty}, {Stahl} and {Zheng}}]{rie22}
\bibinfo{author}{{Riess}, A.G.}, \bibinfo{author}{{Yuan}, W.}, \bibinfo{author}{{Macri}, L.M.}, \bibinfo{author}{{Scolnic}, D.}, \bibinfo{author}{{Brout}, D.}, \bibinfo{author}{{Casertano}, S.}, \bibinfo{author}{{Jones}, D.O.}, \bibinfo{author}{{Murakami}, Y.}, \bibinfo{author}{{Anand}, G.S.}, \bibinfo{author}{{Breuval}, L.}, \bibinfo{author}{{Brink}, T.G.}, \bibinfo{author}{{Filippenko}, A.V.}, \bibinfo{author}{{Hoffmann}, S.}, \bibinfo{author}{{Jha}, S.W.}, \bibinfo{author}{{D'arcy Kenworthy}, W.}, \bibinfo{author}{{Mackenty}, J.}, \bibinfo{author}{{Stahl}, B.E.}, \bibinfo{author}{{Zheng}, W.}, \bibinfo{year}{2022}.
\newblock \bibinfo{title}{{A Comprehensive Measurement of the Local Value of the Hubble Constant with 1 km s$^{-1}$ Mpc$^{-1}$ Uncertainty from the Hubble Space Telescope and the SH0ES Team}}.
\newblock \bibinfo{journal}{\apjl} \bibinfo{volume}{934}, \bibinfo{pages}{L7}.
\newblock \DOIprefix\doi{10.3847/2041-8213/ac5c5b}, \href{http://arxiv.org/abs/2112.04510}{{\tt arXiv:2112.04510}}.
\bibitem[{{Rubin} and {Ford}(1970)}]{1970ApJ...159..379R}
\bibinfo{author}{{Rubin}, V.C.}, \bibinfo{author}{{Ford}, W.~Kent, J.}, \bibinfo{year}{1970}.
\newblock \bibinfo{title}{{Rotation of the Andromeda Nebula from a Spectroscopic Survey of Emission Regions}}.
\newblock \bibinfo{journal}{\apj} \bibinfo{volume}{159}, \bibinfo{pages}{379}.
\newblock \DOIprefix\doi{10.1086/150317}.
\bibitem[{{Salpeter}(1955)}]{sal55}
\bibinfo{author}{{Salpeter}, E.E.}, \bibinfo{year}{1955}.
\newblock \bibinfo{title}{{The Luminosity Function and Stellar Evolution.}}
\newblock \bibinfo{journal}{\apj} \bibinfo{volume}{121}, \bibinfo{pages}{161}.
\newblock \DOIprefix\doi{10.1086/145971}.
\bibitem[{{Sanders} and {McGaugh}(2002)}]{san02}
\bibinfo{author}{{Sanders}, R.H.}, \bibinfo{author}{{McGaugh}, S.S.}, \bibinfo{year}{2002}.
\newblock \bibinfo{title}{{Modified Newtonian Dynamics as an Alternative to Dark Matter}}.
\newblock \bibinfo{journal}{\araa} \bibinfo{volume}{40}, \bibinfo{pages}{263--317}.
\newblock \DOIprefix\doi{10.1146/annurev.astro.40.060401.093923}, \href{http://arxiv.org/abs/astro-ph/0204521}{{\tt arXiv:astro-ph/0204521}}.
\bibitem[{{Slipher}(1915)}]{1915PA.....23...21S}
\bibinfo{author}{{Slipher}, V.M.}, \bibinfo{year}{1915}.
\newblock \bibinfo{title}{{Spectrographic Observations of Nebulae}}.
\newblock \bibinfo{journal}{Popular Astronomy} \bibinfo{volume}{23}, \bibinfo{pages}{21--24}.
\bibitem[{{Stone} et~al.(2021){Stone}, {Courteau} and {Arora}}]{2021ApJ...912...41S}
\bibinfo{author}{{Stone}, C.}, \bibinfo{author}{{Courteau}, S.}, \bibinfo{author}{{Arora}, N.}, \bibinfo{year}{2021}.
\newblock \bibinfo{title}{{The Intrinsic Scatter of Galaxy Scaling Relations}}.
\newblock \bibinfo{journal}{\apj} \bibinfo{volume}{912}, \bibinfo{pages}{41}.
\newblock \DOIprefix\doi{10.3847/1538-4357/abebe4}, \href{http://arxiv.org/abs/2104.07034}{{\tt arXiv:2104.07034}}.
\bibitem[{{Tristram} et~al.(2024){Tristram}, {Banday}, {Douspis}, {Garrido}, {G{\'o}rski}, {Henrot-Versill{\'e}}, {Hergt}, {Ili{\'c}}, {Keskitalo}, {Lagache}, {Lawrence}, {Partridge} and {Scott}}]{tri24}
\bibinfo{author}{{Tristram}, M.}, \bibinfo{author}{{Banday}, A.J.}, \bibinfo{author}{{Douspis}, M.}, \bibinfo{author}{{Garrido}, X.}, \bibinfo{author}{{G{\'o}rski}, K.M.}, \bibinfo{author}{{Henrot-Versill{\'e}}, S.}, \bibinfo{author}{{Hergt}, L.T.}, \bibinfo{author}{{Ili{\'c}}, S.}, \bibinfo{author}{{Keskitalo}, R.}, \bibinfo{author}{{Lagache}, G.}, \bibinfo{author}{{Lawrence}, C.R.}, \bibinfo{author}{{Partridge}, B.}, \bibinfo{author}{{Scott}, D.}, \bibinfo{year}{2024}.
\newblock \bibinfo{title}{{Cosmological parameters derived from the final Planck data release (PR4)}}.
\newblock \bibinfo{journal}{\aap} \bibinfo{volume}{682}, \bibinfo{pages}{A37}.
\newblock \DOIprefix\doi{10.1051/0004-6361/202348015}, \href{http://arxiv.org/abs/2309.10034}{{\tt arXiv:2309.10034}}.
\bibitem[{{Tully} and {Fisher}(1977)}]{tul77}
\bibinfo{author}{{Tully}, R.B.}, \bibinfo{author}{{Fisher}, J.R.}, \bibinfo{year}{1977}.
\newblock \bibinfo{title}{{A new method of determining distances to galaxies.}}
\newblock \bibinfo{journal}{\aap} \bibinfo{volume}{54}, \bibinfo{pages}{661--673}.
\bibitem[{{van Putten}(2017a)}]{van17a}
\bibinfo{author}{{van Putten}, M.H.P.M.}, \bibinfo{year}{2017}a.
\newblock \bibinfo{title}{{Anomalous Galactic Dynamics by Collusion of Rindler and Cosmological Horizons}}.
\newblock \bibinfo{journal}{\apj} \bibinfo{volume}{837}, \bibinfo{pages}{22}.
\newblock \DOIprefix\doi{10.3847/1538-4357/aa5da9}.
\bibitem[{{van Putten}(2017b)}]{van17b}
\bibinfo{author}{{van Putten}, M.H.P.M.}, \bibinfo{year}{2017}b.
\newblock \bibinfo{title}{{Evidence for Galaxy Dynamics Tracing Background Cosmology Below the de Sitter Scale of Acceleration}}.
\newblock \bibinfo{journal}{\apj} \bibinfo{volume}{848}, \bibinfo{pages}{28}.
\newblock \DOIprefix\doi{10.3847/1538-4357/aa88cc}, \href{http://arxiv.org/abs/1709.05944}{{\tt arXiv:1709.05944}}.
\bibitem[{{van Putten}(2018)}]{van18}
\bibinfo{author}{{van Putten}, M.H.P.M.}, \bibinfo{year}{2018}.
\newblock \bibinfo{title}{{Self-similar galaxy dynamics below the de Sitter scale of acceleration}}.
\newblock \bibinfo{journal}{\mnras} \bibinfo{volume}{481}, \bibinfo{pages}{L26--L29}.
\newblock \DOIprefix\doi{10.1093/mnrasl/sly149}, \href{http://arxiv.org/abs/1804.06212}{{\tt arXiv:1804.06212}}.
\bibitem[{{van Putten}(2021)}]{van21}
\bibinfo{author}{{van Putten}, M.H.P.M.}, \bibinfo{year}{2021}.
\newblock \bibinfo{title}{{Evidence of the fine-structure constant in H$_{0}$-tension}}.
\newblock \bibinfo{journal}{Physics Letters B} \bibinfo{volume}{823}, \bibinfo{pages}{136737}.
\newblock \DOIprefix\doi{10.1016/j.physletb.2021.136737}.
\bibitem[{{van Putten}(2024a)}]{van24b}
\bibinfo{author}{{van Putten}, M.H.P.M.}, \bibinfo{year}{2024}a.
\newblock \bibinfo{title}{{Galaxy dynamics tracing quantum cosmology beyond {\ensuremath{\Lambda}}CDM below the de Sitter scale of acceleration}}.
\newblock \bibinfo{journal}{Chinese Journal of Physics} \bibinfo{volume}{91}, \bibinfo{pages}{377--381}.
\newblock \DOIprefix\doi{10.1016/j.cjph.2024.07.040}, \href{http://arxiv.org/abs/2408.06399}{{\tt arXiv:2408.06399}}.
\bibitem[{{van Putten}(2024b)}]{van24}
\bibinfo{author}{{van Putten}, M.H.P.M.}, \bibinfo{year}{2024}b.
\newblock \bibinfo{title}{{The Fast and Furious in JWST high- z galaxies}}.
\newblock \bibinfo{journal}{Physics of the Dark Universe} \bibinfo{volume}{43}, \bibinfo{pages}{101417}.
\newblock \DOIprefix\doi{10.1016/j.dark.2023.101417}, \href{http://arxiv.org/abs/2312.16692}{{\tt arXiv:2312.16692}}.
\bibitem[{{van Putten}(2025)}]{van25}
\bibinfo{author}{{van Putten}, M.H.P.M.}, \bibinfo{year}{2025}.
\newblock \bibinfo{title}{{On the Hubble expansion in a Big Bang quantum cosmology}}.
\newblock \bibinfo{journal}{Journal of High Energy Astrophysics} \bibinfo{volume}{45}, \bibinfo{pages}{194--199}.
\newblock \DOIprefix\doi{10.1016/j.jheap.2024.12.002}, \href{http://arxiv.org/abs/2403.10865}{{\tt arXiv:2403.10865}}.
\bibitem[{{Warner} et~al.(1973){Warner}, {Wright} and {Baldwin}}]{war73}
\bibinfo{author}{{Warner}, P.J.}, \bibinfo{author}{{Wright}, M.C.H.}, \bibinfo{author}{{Baldwin}, J.E.}, \bibinfo{year}{1973}.
\newblock \bibinfo{title}{{High resolution observations of neutral hydrogen in M33 - II. The velocity field.}}
\newblock \bibinfo{journal}{\mnras} \bibinfo{volume}{163}, \bibinfo{pages}{163}.
\newblock \DOIprefix\doi{10.1093/mnras/163.2.163}.
\bibitem[{{Westfall} et~al.(2019){Westfall}, {Cappellari}, {Bershady}, {Bundy}, {Belfiore}, {Ji}, {Law}, {Schaefer}, {Shetty}, {Tremonti}, {Yan}, {Andrews}, {Brownstein}, {Cherinka}, {Coccato}, {Drory}, {Maraston}, {Parikh}, {S{\'a}nchez-Gallego}, {Thomas}, {Weijmans}, {Barrera-Ballesteros}, {Du}, {Goddard}, {Li}, {Masters}, {Ibarra Medel}, {S{\'a}nchez}, {Yang}, {Zheng} and {Zhou}}]{2019AJ....158..231W}
\bibinfo{author}{{Westfall}, K.B.}, \bibinfo{author}{{Cappellari}, M.}, \bibinfo{author}{{Bershady}, M.A.}, \bibinfo{author}{{Bundy}, K.}, \bibinfo{author}{{Belfiore}, F.}, \bibinfo{author}{{Ji}, X.}, \bibinfo{author}{{Law}, D.R.}, \bibinfo{author}{{Schaefer}, A.}, \bibinfo{author}{{Shetty}, S.}, \bibinfo{author}{{Tremonti}, C.A.}, \bibinfo{author}{{Yan}, R.}, \bibinfo{author}{{Andrews}, B.H.}, \bibinfo{author}{{Brownstein}, J.R.}, \bibinfo{author}{{Cherinka}, B.}, \bibinfo{author}{{Coccato}, L.}, \bibinfo{author}{{Drory}, N.}, \bibinfo{author}{{Maraston}, C.}, \bibinfo{author}{{Parikh}, T.}, \bibinfo{author}{{S{\'a}nchez-Gallego}, J.R.}, \bibinfo{author}{{Thomas}, D.}, \bibinfo{author}{{Weijmans}, A.M.}, \bibinfo{author}{{Barrera-Ballesteros}, J.}, \bibinfo{author}{{Du}, C.}, \bibinfo{author}{{Goddard}, D.}, \bibinfo{author}{{Li}, N.}, \bibinfo{author}{{Masters}, K.}, \bibinfo{author}{{Ibarra Medel}, H.J.}, \bibinfo{author}{{S{\'a}nchez}, S.F.}, \bibinfo{author}{{Yang}, M.}, \bibinfo{author}{{Zheng}, Z.},
  \bibinfo{author}{{Zhou}, S.}, \bibinfo{year}{2019}.
\newblock \bibinfo{title}{{The Data Analysis Pipeline for the SDSS-IV MaNGA IFU Galaxy Survey: Overview}}.
\newblock \bibinfo{journal}{\aj} \bibinfo{volume}{158}, \bibinfo{pages}{231}.
\newblock \DOIprefix\doi{10.3847/1538-3881/ab44a2}, \href{http://arxiv.org/abs/1901.00856}{{\tt arXiv:1901.00856}}.
\bibitem[{{Yan} et~al.(2019){Yan}, {Chen}, {Lazarz}, {Bizyaev}, {Maraston}, {Stringfellow}, {McCarthy}, {Meneses-Goytia}, {Law}, {Thomas}, {Falcon Barroso}, {S{\'a}nchez-Gallego}, {Schlafly}, {Zheng}, {Argudo-Fern{\'a}ndez}, {Beaton}, {Beers}, {Bershady}, {Blanton}, {Brownstein}, {Bundy}, {Chambers}, {Cherinka}, {De Lee}, {Drory}, {Galbany}, {Holtzman}, {Imig}, {Kaiser}, {Kinemuchi}, {Liu}, {Luo}, {Magnier}, {Majewski}, {Nair}, {Oravetz}, {Oravetz}, {Pan}, {Sobeck}, {Stassun}, {Talbot}, {Tremonti}, {Waters}, {Weijmans}, {Wilhelm}, {Zasowski}, {Zhao} and {Zhao}}]{2019ApJ...883..175Y}
\bibinfo{author}{{Yan}, R.}, \bibinfo{author}{{Chen}, Y.}, \bibinfo{author}{{Lazarz}, D.}, \bibinfo{author}{{Bizyaev}, D.}, \bibinfo{author}{{Maraston}, C.}, \bibinfo{author}{{Stringfellow}, G.S.}, \bibinfo{author}{{McCarthy}, K.}, \bibinfo{author}{{Meneses-Goytia}, S.}, \bibinfo{author}{{Law}, D.R.}, \bibinfo{author}{{Thomas}, D.}, \bibinfo{author}{{Falcon Barroso}, J.}, \bibinfo{author}{{S{\'a}nchez-Gallego}, J.R.}, \bibinfo{author}{{Schlafly}, E.}, \bibinfo{author}{{Zheng}, Z.}, \bibinfo{author}{{Argudo-Fern{\'a}ndez}, M.}, \bibinfo{author}{{Beaton}, R.L.}, \bibinfo{author}{{Beers}, T.C.}, \bibinfo{author}{{Bershady}, M.}, \bibinfo{author}{{Blanton}, M.R.}, \bibinfo{author}{{Brownstein}, J.}, \bibinfo{author}{{Bundy}, K.}, \bibinfo{author}{{Chambers}, K.C.}, \bibinfo{author}{{Cherinka}, B.}, \bibinfo{author}{{De Lee}, N.}, \bibinfo{author}{{Drory}, N.}, \bibinfo{author}{{Galbany}, L.}, \bibinfo{author}{{Holtzman}, J.}, \bibinfo{author}{{Imig}, J.}, \bibinfo{author}{{Kaiser}, N.},
  \bibinfo{author}{{Kinemuchi}, K.}, \bibinfo{author}{{Liu}, C.}, \bibinfo{author}{{Luo}, A.L.}, \bibinfo{author}{{Magnier}, E.}, \bibinfo{author}{{Majewski}, S.}, \bibinfo{author}{{Nair}, P.}, \bibinfo{author}{{Oravetz}, A.}, \bibinfo{author}{{Oravetz}, D.}, \bibinfo{author}{{Pan}, K.}, \bibinfo{author}{{Sobeck}, J.}, \bibinfo{author}{{Stassun}, K.}, \bibinfo{author}{{Talbot}, M.}, \bibinfo{author}{{Tremonti}, C.}, \bibinfo{author}{{Waters}, C.}, \bibinfo{author}{{Weijmans}, A.M.}, \bibinfo{author}{{Wilhelm}, R.}, \bibinfo{author}{{Zasowski}, G.}, \bibinfo{author}{{Zhao}, G.}, \bibinfo{author}{{Zhao}, Y.H.}, \bibinfo{year}{2019}.
\newblock \bibinfo{title}{{SDSS-IV MaStar: A Large and Comprehensive Empirical Stellar Spectral Library{\textemdash}First Release}}.
\newblock \bibinfo{journal}{\apj} \bibinfo{volume}{883}, \bibinfo{pages}{175}.
\newblock \DOIprefix\doi{10.3847/1538-4357/ab3ebc}, \href{http://arxiv.org/abs/1812.02745}{{\tt arXiv:1812.02745}}.
\bibitem[{{Zhang} et~al.(2024){Zhang}, {Golden-Marx}, {Ogando}, {Yanny}, {Rykoff}, {Allam}, {Aguena}, {Bacon}, {Bocquet}, {Brooks}, {Carnero Rosell}, {Carretero}, {Cheng}, {Conselice}, {Costanzi}, {da Costa}, {Pereira}, {Davis}, {Desai}, {Diehl}, {Doel}, {Ferrero}, {Flaugher}, {Frieman}, {Gruen}, {Gruendl}, {Hinton}, {Hollowood}, {Honscheid}, {James}, {Jeltema}, {Kuehn}, {Kuropatkin}, {Lahav}, {Lee}, {Lima}, {Mena-Fern{\'a}ndez}, {Miquel}, {Palmese}, {Pieres}, {Plazas Malag{\'o}n}, {Romer}, {Sanchez}, {Smith}, {Suchyta}, {Tarle}, {To}, {Tucker}, {Weaverdyck} and {DES Collaboration}}]{zha24}
\bibinfo{author}{{Zhang}, Y.}, \bibinfo{author}{{Golden-Marx}, J.B.}, \bibinfo{author}{{Ogando}, R.L.C.}, \bibinfo{author}{{Yanny}, B.}, \bibinfo{author}{{Rykoff}, E.S.}, \bibinfo{author}{{Allam}, S.}, \bibinfo{author}{{Aguena}, M.}, \bibinfo{author}{{Bacon}, D.}, \bibinfo{author}{{Bocquet}, S.}, \bibinfo{author}{{Brooks}, D.}, \bibinfo{author}{{Carnero Rosell}, A.}, \bibinfo{author}{{Carretero}, J.}, \bibinfo{author}{{Cheng}, T.Y.}, \bibinfo{author}{{Conselice}, C.}, \bibinfo{author}{{Costanzi}, M.}, \bibinfo{author}{{da Costa}, L.N.}, \bibinfo{author}{{Pereira}, M.E.S.}, \bibinfo{author}{{Davis}, T.M.}, \bibinfo{author}{{Desai}, S.}, \bibinfo{author}{{Diehl}, H.T.}, \bibinfo{author}{{Doel}, P.}, \bibinfo{author}{{Ferrero}, I.}, \bibinfo{author}{{Flaugher}, B.}, \bibinfo{author}{{Frieman}, J.}, \bibinfo{author}{{Gruen}, D.}, \bibinfo{author}{{Gruendl}, R.A.}, \bibinfo{author}{{Hinton}, S.R.}, \bibinfo{author}{{Hollowood}, D.L.}, \bibinfo{author}{{Honscheid}, K.}, \bibinfo{author}{{James}, D.J.},
  \bibinfo{author}{{Jeltema}, T.}, \bibinfo{author}{{Kuehn}, K.}, \bibinfo{author}{{Kuropatkin}, N.}, \bibinfo{author}{{Lahav}, O.}, \bibinfo{author}{{Lee}, S.}, \bibinfo{author}{{Lima}, M.}, \bibinfo{author}{{Mena-Fern{\'a}ndez}, J.}, \bibinfo{author}{{Miquel}, R.}, \bibinfo{author}{{Palmese}, A.}, \bibinfo{author}{{Pieres}, A.}, \bibinfo{author}{{Plazas Malag{\'o}n}, A.A.}, \bibinfo{author}{{Romer}, A.K.}, \bibinfo{author}{{Sanchez}, E.}, \bibinfo{author}{{Smith}, M.}, \bibinfo{author}{{Suchyta}, E.}, \bibinfo{author}{{Tarle}, G.}, \bibinfo{author}{{To}, C.}, \bibinfo{author}{{Tucker}, D.L.}, \bibinfo{author}{{Weaverdyck}, N.}, \bibinfo{author}{{DES Collaboration}}, \bibinfo{year}{2024}.
\newblock \bibinfo{title}{{Dark Energy Survey Year 6 results: Intra-cluster light from redshift 0.2 to 0.5}}.
\newblock \bibinfo{journal}{\mnras} \bibinfo{volume}{531}, \bibinfo{pages}{510--529}.
\newblock \DOIprefix\doi{10.1093/mnras/stae1165}, \href{http://arxiv.org/abs/2309.00671}{{\tt arXiv:2309.00671}}.
\bibitem[{{Zibetti} et~al.(2009){Zibetti}, {Charlot} and {Rix}}]{Zibetti09}
\bibinfo{author}{{Zibetti}, S.}, \bibinfo{author}{{Charlot}, S.}, \bibinfo{author}{{Rix}, H.W.}, \bibinfo{year}{2009}.
\newblock \bibinfo{title}{{Resolved stellar mass maps of galaxies - I. Method and implications for global mass estimates}}.
\newblock \bibinfo{journal}{\mnras} \bibinfo{volume}{400}, \bibinfo{pages}{1181--1198}.
\newblock \DOIprefix\doi{10.1111/j.1365-2966.2009.15528.x}, \href{http://arxiv.org/abs/0904.4252}{{\tt arXiv:0904.4252}}.
\bibitem[{{Zwicky}(1937)}]{1937ApJ....86..217Z}
\bibinfo{author}{{Zwicky}, F.}, \bibinfo{year}{1937}.
\newblock \bibinfo{title}{{On the Masses of Nebulae and of Clusters of Nebulae}}.
\newblock \bibinfo{journal}{\apj} \bibinfo{volume}{86}, \bibinfo{pages}{217}.
\newblock \DOIprefix\doi{10.1086/143864}.

\end{thebibliography}

\mbox{}\\
{\bf SUPPORTING INFORMATION}\\
Supplementary data on $\Delta_{b}{\rm PA}$ are available online.

\label{lastpage}
\end{document}